\documentclass[11pt]{article}
\usepackage{authblk}
\pdfoutput=1
\usepackage[style=numeric-comp,maxnames=9,sorting=none,backend=biber]{biblatex}
\usepackage{pslatex}    
\usepackage{xcolor}
\usepackage{amsfonts}
\usepackage{amsmath}
\usepackage{amsthm}
\usepackage{amscd}
\usepackage{eucal}
\usepackage{graphicx}
\usepackage[a4paper, total={7in, 9.5in}, top=1.2in]{geometry}
\usepackage{manfnt}
\usepackage{booktabs}
\usepackage{microtype}
\usepackage{soul}
\usepackage{csquotes}

\AtEveryBibitem{%
  \clearfield{note}%
  \clearfield{reportNumber}%
}

\newcommand{\nc}{\newcommand}
\nc{\rnc}{\renewcommand}
\nc{\non}{\nonumber}

\nc{\ir}{\mathrm{i}}
\def\er{{\mathrm e}}

\def\hToo{\hat\Lambda_I}
\def\hTot{\hat\Lambda_{II}}
\def\BAnum{N}
\def\Rmat{{\mathrm R}}
\def\beq{\begin{equation}}
\def\eeq{\end{equation}}
\def\bea{\begin{eqnarray}}
\def\eea{\end{eqnarray}}

\def\beann{\begin{eqnarray*}}
\def\eeann{\end{eqnarray*}}

\let\thet=\theta

\theoremstyle{plain}

\newtheorem*{corollary*}{Corollary}

\newtheorem*{conjecture*}{Conjecture}

\theoremstyle{definition}

\def\2{\frac{1}{2}} \def\4{\frac{1}{4}}

\def\6{\partial}

\def\+{\dagger}

\def\<{\langle} \def\>{\rangle}

\def\i{{\rm i}}

\def\rd{{\rm d}}
\def\re{{\rm e}}

\def\Re{{\rm Re\,}} \def\Im{{\rm Im\,}}

\def\av{{a}}

\def\Av{{A}}

\def\Ff{\mathfrak{F}}

\def\jg2{K}

\renewcommand{\tilde}{\widetilde}
\begin{document}
\title{Managing Singular Kernels and Logarithmic Corrections in the Staggered
Six-Vertex Model}

\bigskip
\author[1,2]{Mouhcine Azhari}
\author[1]{Andreas Kl\"{u}mper}
\affil[1]{Department of Physics,
 University of Wuppertal, Gaussstra\ss e 20, 42119 Wuppertal, Germany}
\affil[2]{\'Ecole Royale Navale, Casablanca 20052, Morocco}
\affil[ ]{\texttt{azhari.mouhcine@gmail.com}, \texttt{kluemper@uni-wuppertal.de}}

\maketitle

\begin{abstract}
In this paper, we investigate the spectral properties of the staggered
six-vertex model with ${\cal Z}_2$ symmetry for arbitrary
system sizes $L$ using non-linear integral equations (NLIEs). Our study is
motivated by two key questions: what is the accuracy of results based on the ODE/IQFT
correspondence in the asymptotic regime of large system sizes, and what is the optimal
approach based on NLIE for analyzing the staggered six-vertex model?

We demonstrate that the quantization conditions for low-lying primary and
descendant states, derived from the ODE/IQFT approach in the scaling limit, are impressively
accurate even for relatively small system sizes. Specifically, in the
anisotropy parameter range $\pi/4 < \gamma < \pi/2$, the difference between
NLIE and ODE/IQFT results for energy and quasi-momentum eigenvalues 
is of order $\mathcal{O}(L^{-2})$.

Furthermore, we present a unifying framework for NLIEs, distinguishing between
versions with singular and regular kernels. We provide a compact derivation of
NLIE with a singular kernel, followed by an equivalent set with a regular
kernel. We address the stability issues in numerical treatments and offer
solutions to achieve high-accuracy results, validating our approach for system
sizes ranging from $L=2$ to $L=10^{24}$.

Our findings not only validate the ODE/IQFT approach for finite system sizes
but also enhance the understanding of NLIEs in the context of the staggered
six-vertex model. We hope the insights gained from this study have significant
implications for resolving the spectral problem of other lattice systems with
emergent non-compact degrees of freedom and provide a foundation for future
research in this domain.

\end{abstract}



\section{Introduction}

The study of two-dimensional (2D) integrable lattice systems, such as the
Ising model, the six-vertex model, and the eight-vertex model, alongside their
associated one-dimensional (1D) integrable quantum spin chains, has
significantly shaped our understanding of universality and critical
phenomena. These models continue to be pivotal in the exploration of
correlation functions both in and out of equilibrium
\cite{jimbo2021local,boos2010local,vasseur2016nonequilibrium}, and in the
study of quantum entanglement \cite{bayat2022entanglement}.

A basic yet fundamental example is the homogeneous six-vertex model,
particularly when the anisotropy parameter $q$ is unimodular ($|q| = 1$). In
the scaling limit, this model is described by a compact massless 1+1
dimensional boson field. However, more intriguing behaviors emerge in the
integrable, inhomogeneous six-vertex model introduced by Baxter
\cite{baxter1971eight}. Specifically, when staggered inhomogeneities are
introduced, the model exhibits two distinct types of universal behavior
depending on the value of $q$ \cite{jacobsen2006critical}.

Jacobsen and Saleur \cite{jacobsen2006critical,ikhlef2008staggered} revealed
that the theory possesses a continuous spectrum of scaling dimensions. A
significant conjecture by Ikhlef, Jacobsen, and Saleur
\cite{ikhlef2012conformal} proposed that the staggered lattice model realizes
the Euclidean black hole non-linear sigma model (NLSM) in the scaling limit,
connecting it to works on black hole CFTs
\cite{elitzur1991some,mandal1991,witten1991string,dijkgraaf1992,maldacena2001strings,hanany2002partition}. This
conjecture was confirmed and further elaborated by Bazhanov, Kotousov, Koval,
and Lukyanov \cite{bazhanov2019scaling,bazhanov2021scaling} and
\cite{bazhanov2021equilibrium} through the ODE/IQFT approach, establishing the
scaling limit is related to the 2D Euclidean/Lorentzian black hole CFTs. The
lattice analysis has been essential in resolving the spectral problem for the
2D Euclidean black hole field theory, including determining the density of
states for the continuous spectrum \cite{bazhanov2021scaling}.

Linear integral equations for density functions of Bethe roots for models with
emerging non-compact degrees of freedom have been used by Ikhlef, Jacobsen,
and Saleur \cite{ikhlef2010staggered,ikhlef2012conformal}, as well as by Frahm
and Martins \cite{frahm2011finite,frahm2012phase}. These equations are derived
for density functions defined in the thermodynamic limit. Results obtained
from these density functions, calculated either numerically or, under certain
conditions, analytically, yield bulk properties. When finite ``Fermi points"
are considered, this approach allows for Sommerfeld-like computations of
finite size quantities. In cases where the distributions of roots span the
entire real axis, the Wiener-Hopf technique may be applied
\cite{ikhlef2012conformal}. Our paper addresses the case of strictly finite
size systems.

For finite systems, non-linear integral equations (NLIEs) based on
the Bethe Ansatz have been instrumental. Candu and Ikhlef \cite{Candu_2013}
as well as Frahm and Seel \cite{Frahm14}, were the first to
derive NLIEs for the staggered six-vertex model, significantly contributing to
the understanding and solution of these equations. In this paper, we present a
unifying framework for NLIEs, particularly useful for high-accuracy
calculations in the context of the staggered six-vertex model.

We study the spectral properties of the staggered six-vertex model with
${\cal Z}_2$ symmetry. We are primarily interested in the accuracy of the
results based on the ODE/IQFT correspondence in the asymptotic regime of large
system sizes, and in finding the optimal approach based on NLIE for studying
the staggered six-vertex model. We found that the quantization conditions for
low-lying primary states obtained in \cite{ikhlef2012conformal} and improved
and extended to descendant states in the works of Bazhanov et al.
\cite{bazhanov2019scaling,bazhanov2021scaling,bazhanov2021some,bazhanov2021equilibrium},
are impressively accurate for a wide range of sizes. In the anisotropy
parameter range $\pi/4 < \gamma < \pi/2$, the difference between NLIE and
ODE/IQFT results for energy and quasi-momentum eigenvalues vanishes in the
scaling limit as $\mathcal{O}(L^{-2})$.

We also analyze different versions of NLIE, noting that while some have
singular kernels, others have regular kernels. We present a compact derivation
of the NLIE with singular kernel from which the relationship to the other
versions is understood. By rearranging terms, we derive an equivalent set of
NLIE with regular kernel. We explain the stability issues in the numerical iterative
treatment of these equations and provide solutions with high numerical
accuracy. We performed calculations for lattice sizes ranging from $L=2$ to
$L=10^{24}$, showing excellent agreement with ODE/IQFT results.

The structure of this paper is as follows: Section 2 reviews the fundamental
aspects of the six-vertex model and its staggered variant. Section 3 presents
the derivation of our first version of NLIE with a singular integral kernel
for the lowest-lying states of the model. In Section 4, we derive an
equivalent version with a regular kernel and present numerical results for
various system sizes, comparing these results to those obtained using the
ODE/IQFT approach. Section 5 presents the analytic derivation of the main
logarithmic term in the spectrum using the singular kernel version of our
NLIEs without resorting to Wiener-Hopf techniques. Section 6 addresses
necessary modifications of the NLIE for dealing with non-primary
states. Finally, in Section 7, we summarize our results, suggest directions
for future research, and provide more technical information in the Appendix.

\section{The staggered six-vertex model: definition and eigenvalue equation}
We study the six-vertex model with standard $\Rmat$-matrix acting on the
tensor product of two 2-dimensional vector spaces by following in this section
largely the notation of \cite{Candu_2013}. The only non-zero elements
of $\Rmat$ in the standard basis, $| + + \rangle, | + -\rangle, | - +
\rangle, | - - \rangle$, are
\begin{equation}
\Rmat^{++}_{++}(u)=\Rmat^{--}_{--}(u)=a(u)\,,\quad 
\Rmat^{+-}_{+-}(u)=\Rmat^{-+}_{-+}(u)=
b(u)\,,\quad
\Rmat^{+-}_{-+}(u)=\Rmat^{-+}_{+-}(u)=c(u)\,.\label{rmat}
\end{equation}
where $a(u)={\sin(\gamma-u)}$, $b(u)={\sin(u)}$, $c(u)={\sin(\gamma)}$, 
$u$ is referred to as the spectral parameter, and $\gamma$ is the anisotropy parameter.  A
suitable product of $\Rmat$-matrices with same spectral parameter $u$
yields the commuting family of row-to-row transfer matrices $T(u)$ whose
logarithmic derivative at $u=0$ is the Hamiltonian of the $S=1/2$ $XXZ$
quantum spin chain. In our paper we use periodic boundary conditions.

The model allows for generalizations keeping its integrability
\cite{baxter1971eight}. A modest generalization is given by the staggering of
the spectral parameter from colum to column \cite{jacobsen2006critical}. The
alternating product of $\Rmat(u)$ and $\Rmat(u-\alpha)$, with fixed parameter
$\alpha$, in dependence on $u$ still yields a commuting family of transfer
matrices, which for simplicity again will be called $T(u)$.

\begin{figure}
\begin{center}
\includegraphics[scale=0.9]{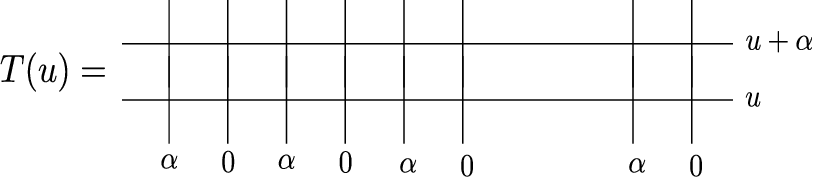}
\caption{Depiction of the double row transfer matrix of the staggered 6-vertex model.}
\label{fig:block-Rmat}
\end{center}
\end{figure}

The logarithmic derivative of $T(u)$ at $u=0$ resp.~at $u=\alpha$
yields a sum
of locally acting terms, unfortunately without particular symmetries. However,
the sum of these operators, i.e.~the logarithmic derivative of the product
$T(u)T(u+\alpha)$ at $u=0$ exhibits $\cal{CP}$ and $\cal{T}$ invariance
\cite{bazhanov2021scaling}. This follows among other things from unitarity and
the standard initial condition of the
$\Rmat$-matrix. Fig.~\ref{fig:block-Rmat} shows an illustration of the
construction principle on the square lattice. To each vertical line a spectral
parameter 0 or $\alpha$ is assigned and to the horizontal lines $u$ or
$u+\alpha$.  Each vertex is associated with an $\Rmat$-matrix whose argument
is given by the difference of the spectral parameters on the intersecting
horizontal and vertical lines. The lower row corresponds to $T(u)$, the upper
row to $T(u+\alpha)$.
For general values of $\alpha$ the double row transfer matrix with periodic
boundary conditions is invariant under a translation by 2 lattice sites. For
$\alpha=\pi/2$ an invariance under a translation by 1 lattice site is
restored, and also, the system shows a ${\cal Z}_2$ symmetry
\cite{bazhanov2021scaling}. Note that generalizations of staggered six-vertex
models with ${\cal Z}_r$ symmetry have been studied in
\cite{kotousov2023scaling,bazhanov2021some}.

The quantum Hamiltonian of the spin-$\frac{1}{2}$ chain of length $2L$
associated with the double-row transfer matrix is expressed in terms of Pauli
matrices which for the case $\alpha=\pi/2$ has explicit form 
\begin{align}
  H&=\frac12\sin(2\gamma)\cdot\partial_u\log\left[T(u)T(u+\pi/2)\right]\big|_{u=0}\cr
  &=\sum_{j=1}^{2L}\left[-\frac12{\vec\sigma}_j{\vec\sigma}_{j+2}+\sin^2(\gamma)\,
    \sigma_j^z\sigma_{j+1}^z-\frac{\ir}2\sin(\gamma)\left(\sigma_{j-1}^z-\sigma_{j+2}^z\right)\left(\sigma_{j}^x\sigma_{j+1}^x+\sigma_{j}^y\sigma_{j+1}^y\right)\right]+L\cos(2\gamma).
\end{align}
In this paper, we restrict ourselves to the regime of the anisotropy parameter
${\pi}/{4}<\gamma<{\pi}/{2}$.

As is well known, each eigenvalue $\Lambda$ of the transfer matrix $T(u)$ of the
six-vertex model satisfies the $T$-$q$ equation. This reads
\begin{equation}
    \Lambda(z) q(z) = \Phi(z-\ir\gamma) q(z+2\ir\gamma) + \Phi(z+\ir\gamma) q(z-2\ir\gamma),\label{Gleichung}
\end{equation}
after a transformation of the variable $u=-(\i/2)z+\gamma/2$ to the
argument $z$.
The factor $\Phi(z-\ir\gamma)$ is the $L$-th power of
$a(u)a(u-\pi/2)=\sin(\gamma-u)\sin(\gamma-u+\pi/2)=(\i/2)\sinh(z-\i\gamma)$,
and $\Phi(z+\ir\gamma)$ is the $L$-th power of
$b(u)b(u-\pi/2)=\sin(u)\sin(u-\pi/2)=(\i/2)\sinh(z+\i\gamma)$.
Hence
\begin{equation}\label{eq:4}
 \Phi(z):=\sinh^L(z),\qquad  q(z):=\prod_{j=1}^{\BAnum}\sinh \tfrac12(z-z_j),
\end{equation}
where we have dropped the factor $(\i/2)^L$ in $\Phi$. Whenever necessary this
factor has to be retrieved. However, in the following we will be interested in
logarithmic derivatives and ratios of the eigenvalue functions.

\begin{figure}
\begin{center}
\includegraphics[scale=0.4]{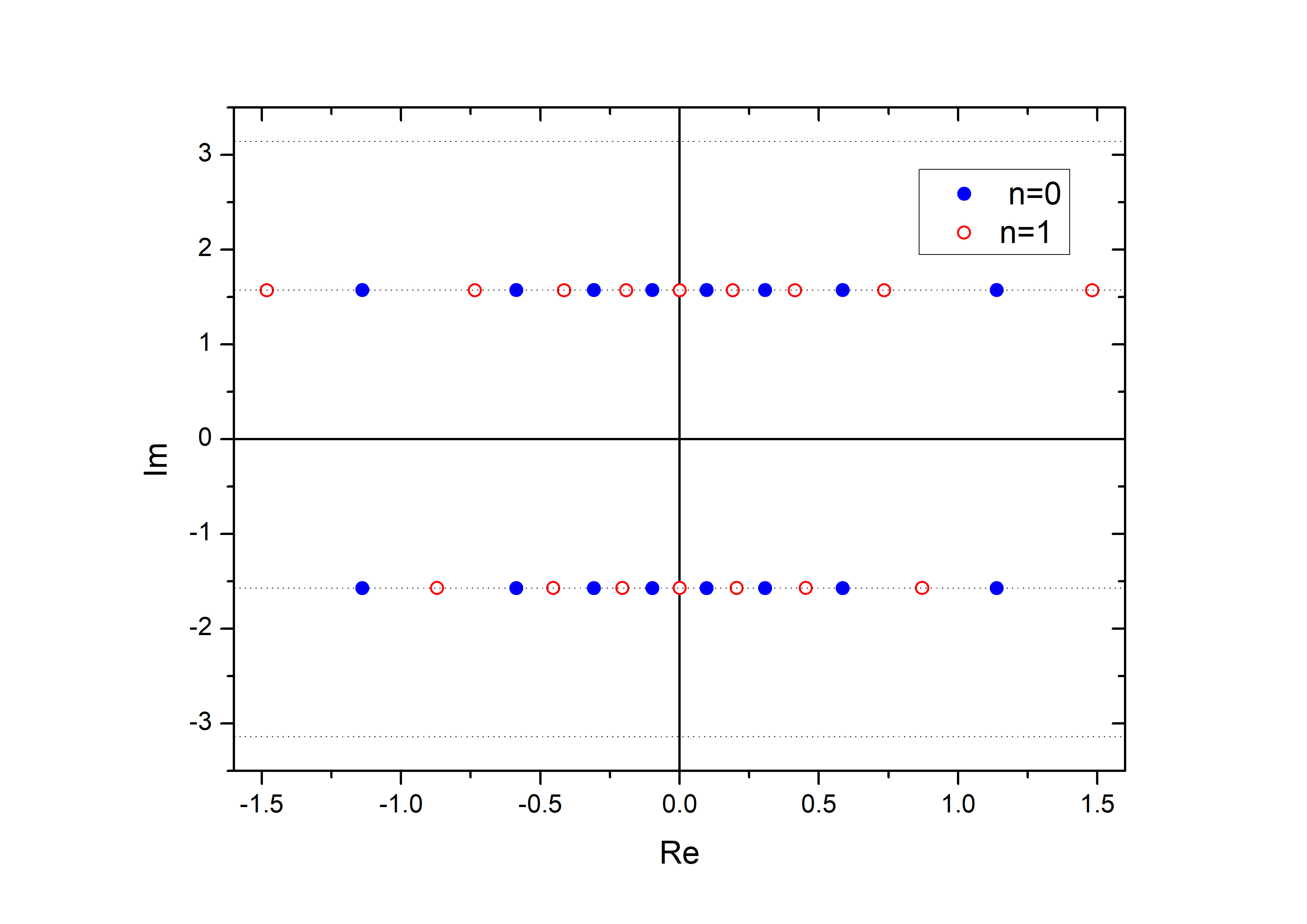}  
  \caption{Depiction of the location of all zeros of $q(z)$ in the complex
    plane for $L=16$ and $\gamma=0.8$. Disks refer to the ground state ($n=0$)
    and circles to the first excited state ($n=1$).}
\label{fig:BArap}
\end{center}
\end{figure}

The Bethe ansatz roots (or rapidities) $z_j$, $j=1, \ldots, \BAnum$, $0 \le \BAnum \le L$,
have to satisfy the Bethe ansatz equations:
\begin{equation}
    \av(z_j)=-1,\qquad j=1, \ldots, \BAnum,
\end{equation}
where the function $\av(z)$ -- not to be confused with the Boltzmann weight
appearing in (\ref{rmat}) -- is defined as 
\begin{equation}\label{eq:6}
    \av(z):=\frac{\Phi(z+\ir \gamma)q(z-2\ir \gamma)}{\Phi(z-\ir \gamma)q(z+2\ir \gamma)}.
\end{equation}
For the largest eigenvalue and a large class of next-leading eigenvalues the
total number of roots is equal to $L$ with the distribution of the roots along
the horizontal axes through $\pm \ir{\pi}/{2}$. Here we consider $L/2\pm n$
roots located on the upper/lower axis with $L$ typically much
larger than $n$.  See Fig.~\ref{fig:BArap}
for anisotropy parameter
$\gamma=0.8$ for distributions of roots for $L=16$ with $n=0$ and $1$.
%
%
%
The truely largest eigenvalue $\Lambda$, corresponding to the ground state
energy of the quantum chain, is given by the case $n=0$.

In the next sections we restrict ourselves to the low-lying states just
described. The CFT descendants to these states are considered in
Sect.~\ref{sect:Desc}.  More general cases, primary states corresponding to
distributions of roots with holes and descendants to these, will be published
in separate work.

\section{Non-linear integral equations: singular kernel version}\label{sect:NLIEsing}

The eigenvalues $\Lambda(z)$ of $T(z)$ are analytic functions of the spectral
parameter $u$. In this paper, we use this analyticity and adopt
a method in analogy to the approach in~\cite{Klumper91}, see also
\cite{klumper1990analytic,klumper1992free,DeVe92,DesVeg95,suzuki1999spinons},
to determine the eigenvalues by a finite set of non-linear integral equations
(NLIE) and valid for arbitrary (even) system size. This
approach also allows for taking the limit $L\rightarrow \infty$
analytically. The central object is the function $\av(z)$ (\ref{eq:6}) in the
complex plane. It has zeros of order $L$ at $-\ir\gamma$, $(\pi-\gamma)\ir$
and poles of order $L$ at $+\ir\gamma$, $(\pi+\gamma)\ir$. As we will see
shortly, it will be useful to consider $\av(z)$ and its reciprocal on the
horizontal axes through the points of high order zeros and poles:
\begin{align}
a_1(x)&:=\frac1{a(x+\ir\gamma)}=\frac{\Phi(x)}{\Phi(x+2\ir\gamma)}\frac{q(x+3\ir\gamma)}{q(x-\ir\gamma)},\label{deflittlea1}\\
a_2(x)&:={a(x+\ir\pi-\ir\gamma)}=\frac{\Phi(x)}{\Phi(x-2\ir\gamma)}\frac{q(x+\ir\pi-3\ir\gamma)}{q(x+\ir\pi+\ir\gamma)},\\
a_3(x)&:={a(x-\ir\gamma)}=\frac{\Phi(x)}{\Phi(x-2\ir\gamma)}\frac{q(x-3\ir\gamma)}{q(x+\ir\gamma)},\\
a_4(x)&:=\frac1{a(x+\ir\pi+\ir\gamma)}=\frac{\Phi(x)}{\Phi(x+2\ir\gamma)}\frac{q(x+\ir\pi+3\ir\gamma)}{q(x+\ir\pi-\ir\gamma)},\label{deflittlea4}
\end{align}
where the explicit factorization in terms of $\Phi(z)$ and $q(z)$ is shown on
the right.  These functions by themselves do not define a closed system of
functional equations. This is achieved by introducing the additional set of
closely related functions
\begin{equation}
\Av_i(z):=1+\av_i(z).
\end{equation}  
By use of the functional equation (\ref{Gleichung}) these functions
can be written in terms of $\Lambda(z)$ as follows
\begin{align}
  A_1(x)&=\frac{1}{\Phi(x+2\ir\gamma)}
  \frac{q(x+\ir\gamma)}{q(x-\ir\gamma)}\Lambda(x+\ir\gamma)\label{defbigA1},\\
  A_2(x)&=\frac{1}{\Phi(x-2\ir\gamma)}
  \frac{q(x+\ir\pi-\ir\gamma)}{q(x+\ir\pi+\ir\gamma)}\Lambda(x+\ir\pi-\ir\gamma)\label{defbigA2},\\
  A_3(x)&=\frac{1}{\Phi(x-2\ir\gamma)}
  \frac{q(x-\ir\gamma)}{q(x+\ir\gamma)}\Lambda(x-\ir\gamma)\label{defbigA3},\\
  A_4(x)&=\frac{1}{\Phi(x+2\ir\gamma)}
  \frac{q(x+\ir\pi+\ir\gamma)}{q(x+\ir\pi-\ir\gamma)}\Lambda(x+\ir\pi+\ir\gamma).\label{defbigA4}
\end{align}



The key of our approach is the identification of analyticity domains in which
the functions $q(z)$ and $\Lambda(z)$ are free of zeros, i.e. domains where
these functions are analytic and non-zero (ANZ). We find two such analyticity
strips for each function modulo $2\pi\ir$-periodicity. Hence there exist four
Fourier transform expressions for all $q$ and $\Lambda$ data on the right hand
sides of (\ref{deflittlea1})-(\ref{deflittlea4}) and
(\ref{defbigA1})-(\ref{defbigA4}). This will allow for the derivation of a
closed set of integral equations for the functions $a_i$ and for integral
expressions for the eigenvalue function.

\begin{figure}
\begin{center}
\includegraphics[scale=0.3]{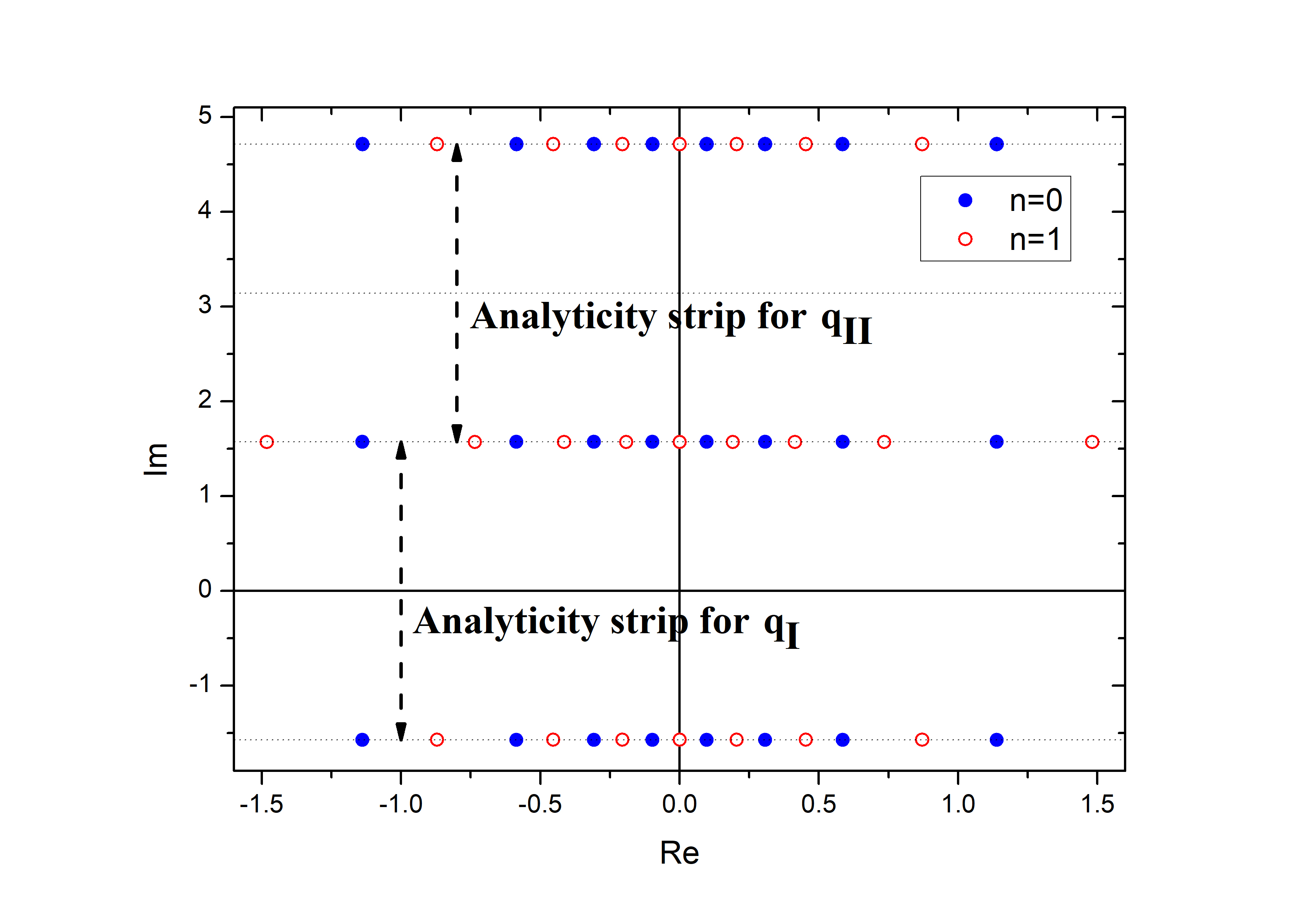}
\includegraphics[scale=0.3]{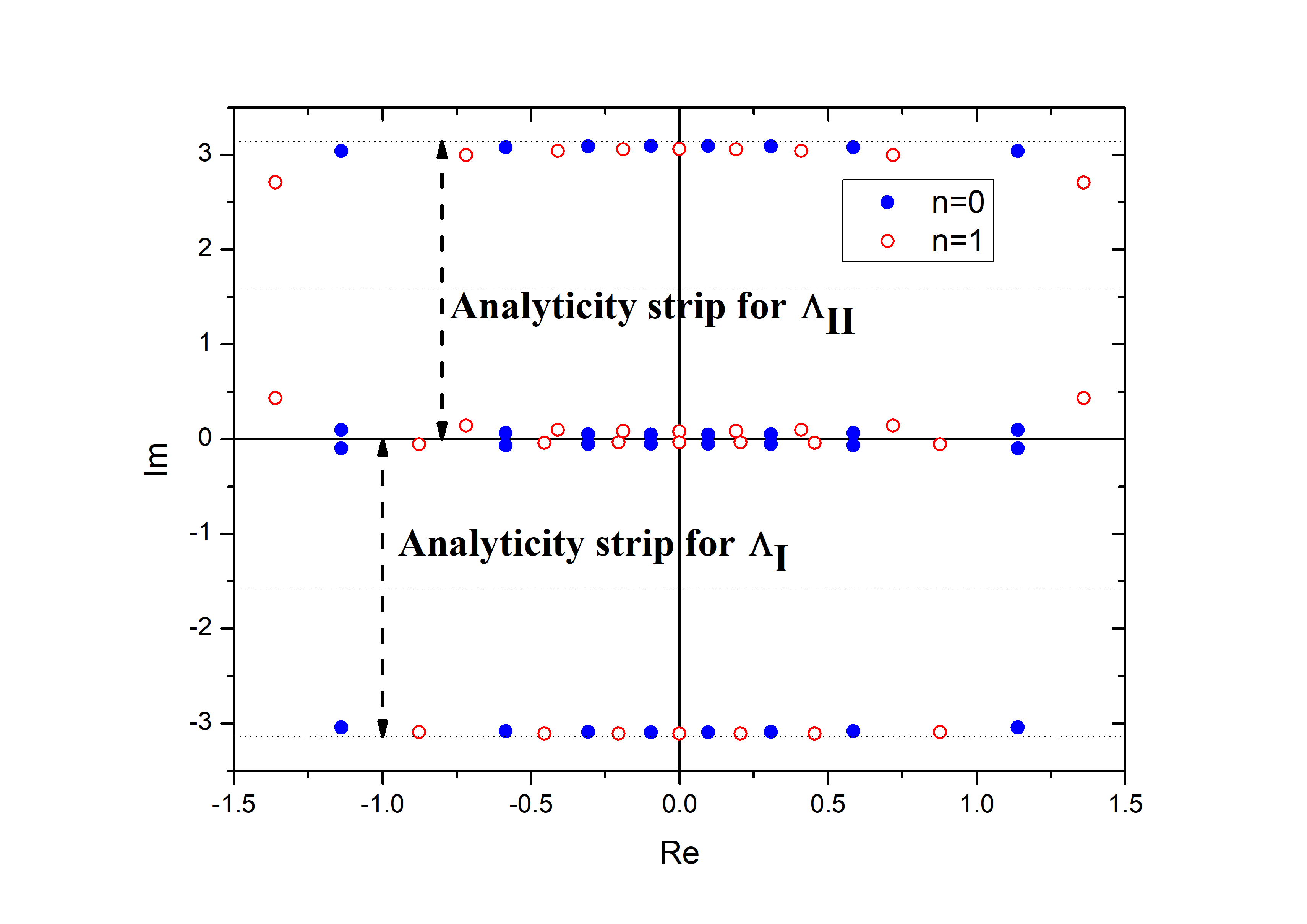}  
\caption{Depiction of the location of all zeros in the complex plane for a)
  the function
  $q(z)$ and b) the eigenvalue function $\Lambda(z)$. We use parameters $L=16$
  and $\gamma=0.8$. Note that the functions are $2\pi\i$ periodic.  Disks
  refer to the ground state ($n=0$) and circles to the first excited state
  ($n=1$). The zeros of $q(z)$ lie precisely on straight lines, the zeros of
  $\Lambda(z)$ are close to $\mathbb{R}$ and to $\mathbb{R}\pm \ir{\pi}$, but
  deviate from them. The strips of the complex plane that are free of zeros of
  $q(z)$ and $\Lambda(z)$ allow for the Fourier transform of the logarithmic
  derivatives of the functions. We refer to these Fourier representations in
  different strips by indices I and II to indicate which
  analyticity strip is used. Note that strips I and II are different
  for $q(z)$ and $\Lambda(z)$.}
\label{fig:zeros-q-Lambda}
\end{center}
\end{figure}

By periodicity we can restrict our attention to fundamental regions of height
$2\pi$ in the complex plane. These are taken slightly differently for the
functions $q(z)$ and $\Lambda(z)$: $-\tfrac12\pi<\Im(z)<\tfrac32\pi$ and
$0<\Im(z)<2\pi$, see Fig.~\ref{fig:zeros-q-Lambda} for parameters $L=16$ and
$\gamma=0.8$. In the same figure, the zeros of $\Lambda(z)$ are also shown. An
index I resp.~II labels the strips where no zeros are present.  The ANZ
strips for $q(z)$ are (I) $-\tfrac12\pi<\Im(z)<\tfrac12\pi$ and (II)
$\tfrac12\pi<\Im(z)<\tfrac32\pi$. For $\Lambda(z)$ the ANZ strips are (I) 
$-\pi<\Im(z)<0$ and (II) $0<\Im(z)<\pi$ (actually a little narrower due to the
deviations of zeros from the straight lines).

First we observe from (\ref{deflittlea1})-(\ref{deflittlea4}) and
(\ref{defbigA1})-(\ref{defbigA4})
that the functions $\av_i$ take very small values for argument $x$ in the
neighbourhood of $0$ and have simple asymptotes
\begin{equation}
    \av_i(\pm \infty)=1,\qquad  \Av_i(\pm \infty)=1+\av_i(\pm \infty)=2,\qquad i=1, \ldots, 4.
\end{equation}
Their continuous logarithms for argument $x$ along the real axis have well
defined finite asymptotic values for $-\infty$ and $+\infty$. Note that
in general these asymptotic values need not be the same. Notably for the
functions $\log a_i(x)$ we observe an increase or decrease of the imaginary
part by multiples of $2\pi$. This value, i.e.~the winding of $a_i(x)$ around
$0$ for $x$ from $-\infty$ to $+\infty$, serves as a strict quantization
condition for the eigenvalues. Of course, the logarithmic derivative of all functions
can be Fourier-transformed and equations (\ref{deflittlea1})-(\ref{deflittlea4}) and
(\ref{defbigA1})-(\ref{defbigA4}) turn into eight linear relations for the
Fourier transforms. We like to note that the logarithmic derivatives of $q$
and $\Lambda$ have non-zero and different asymptotic values for the real part
of the argument to
$-\infty$ and $+\infty$. Therefore strictly speaking we should work with
Fourier transforms of second derivatives of the logarithms of all involved
functions. 

Next we fix the definitions of the Fourier transform $\Ff_{k}\{f\}$ (and its
inverse) of a complex function $f(z)$ which is analytic in a certain strip and
decays sufficiently fast. The Fourier transform pair is
\begin{equation}
    \Ff_{k}\{f\}:=\frac{1}{2\pi}\int_{{\cal C}} f(x) \er^{-\ir kx}\rd x,\qquad
    f(x)=\int_{-\infty}^{\infty} \Ff_{k}\{f\} \er^{\ir k x}\rd k.
\end{equation}
where the integration path ${\cal C}$ is the real axis or has to lie in the
appropriate analyticity strip of the functions $q(z)$ and $\Lambda(z)$. Note
that the Fourier transforms of (the derivative of) $\log q(z)$ in strips I and
II differ. For that reason we index these functions by I and II and have to
consider them as different. The same remark applies to the function
$\Lambda(z)$. Note that the inverse Fourier transforms with $k$-integrals along
the real axis have different convergence regions for different indices I and
II: these are the above identified ANZ strips I and II.

For multiplicative relations like
\begin{equation*}
f(x)=g(x+\ir\alpha) / h(x+\ir\beta),
\end{equation*}
with shifts leaving the argument in the analyticity strip of the respective function,
the Fourier transform of the logarithmic derivative yields
\begin{equation*}
\Ff_{k}\{(\log f)'\}=
\er^{-\alpha k}\,\Ff_{k}\{(\log g)'\}-\er^{-\beta k}\,\Ff_{k}\{(\log h)'\}.
\end{equation*}
This relation also holds for the Fourier transform of the second logarithmic derivative.

The eight multiplicative relations (\ref{deflittlea1})-(\ref{deflittlea4}) and
(\ref{defbigA1})-(\ref{defbigA4}) yield eight linear equations for the Fourier
transforms of the logarithmic derivatives of the functions
$\av_1, \av_2, \av_3, \av_4, \Av_1, \Av_2, \Av_3, \Av_4, q_I, q_{II},
\Lambda_I$ and $\Lambda_{II}$. These can be solved uniquely for
$\av_1, \av_2, \av_3, \av_4, q_I, q_{II}, \Lambda_I$ and $\Lambda_{II}$ in
terms of $\Av_1,\Av_2, \Av_3,\Av_4$ with certain $k$-dependent factors. The
inverse Fourier transform consists of explicit functions and in addition of
convolution integrals of explicit functions with $\log \Av_i$
functions. Details can be found in the appendix.

We apply this strategy, do the inverse Fourier transform and obtain integral
equations first for the differentiated logarithms of $\av_i$ in terms of those
for $\Av_i$. We finally integrate with respect to $x$, identify the
integration constants and obtain, the following compact form of the NLIE of
the staggered 6-vertex model:
\begin{equation}
\av
  =d+K\ast \Av,\label{singNLIE}
  \end{equation}
 where
\begin{equation}
  \av(x)=
    \left(\begin{matrix}
    \log \av_1(x)\\
    \log \av_2(x)\\
    \log \av_3(x)\\
    \log \av_4(x)
  \end{matrix}\right),
  \quad
  \Av(x)=
  \left(\begin{matrix}
    \log \Av_1(x)\\
    \log \Av_2(x)\\
    \log \Av_3(x)\\
    \log \Av_4(x)
  \end{matrix}\right),
  \quad
  d(x)= L\,\log(\tanh(\tfrac12gx))\cdot
  \left(\begin{matrix}
    1\\
    1\\
    1\\
    1
  \end{matrix}\right),\quad
  g:=\frac{\pi}{\pi-2\gamma},\label{abbreviations}
  \end{equation}
where $d(x)$ is the driving resp. source term. The notation $g*f$ denotes
the convolution of the functions $g$ and $f$,
\begin{equation}
    (g\ast f)(x)=\frac1{2\pi}\int_{-\infty}^{\infty} g(x-y)f(y) \rd y,
\end{equation}
where we introduced the prefactor $1/2\pi$ for convenience in some situations.
This set of integral equations holds for any imbalance $n$ of the number of
roots on the axes with imaginary parts $\pm\pi/2$.

The kernel has the following block structure
\begin{equation}
    K=\left(\begin{matrix}
\sigma_1  &\sigma_2\cr
\sigma_2^\dagger & \sigma_1^T
\end{matrix}\right),\qquad\qquad (\dagger\ \hbox{interchanges diagonal
      elements}).
\label{FTKernelInv}    
\end{equation}
where in Fourier representation we have
\begin{align}
    \sigma_1&=
\frac{\cosh((\pi-3\gamma)k)}{2\sinh(\gamma k)\sinh((\pi-2\gamma)k)}
\left(\begin{matrix}
-1  &\er^{(\pi-2\gamma)k}\cr
\er^{(2\gamma-\pi)k}& -1
\end{matrix}\right),\label{block1}\\
\sigma_2&=
\frac{\cosh(\gamma k)}{2\sinh(\gamma k)\sinh((\pi-2\gamma)k)}
\left(\begin{matrix}
-\er^{(\pi-2\gamma)k} & 1\cr
1& -\er^{(2\gamma-\pi)k}
\end{matrix}\right).\label{block2}
\end{align}
The kernel is singular as it has a pole of second order at $k=0$.  The inverse
Fourier transform involves functions that do not decay at large distances,
they increase linearly!

At first glance these integral equations look useless and possibly
ill-defined. The right way to look at them is that the left-hand side is well
defined and via the integral equations imposes a condition on the functions on
the right-hand side in order for the convolution integral to exist. This applies to
the solutions to the integral equations. Unfortunately, it does not apply to
all functions generated in an iterative application of the equations when
starting with some initial data.

Of course, when using the above set of integral equations we have to say how
the inverse Fourier transform of (\ref{FTKernelInv}) is precisely
defined. This we treat in section \ref{Sect:AnalyticalStudy} and in the
appendix. In section \ref{sect:NLIEreg}
we will derive from (\ref{singNLIE}) an alternative
equivalent set of NLIE however with regular kernel. Before doing so we like to
present some results to illustrate some properties of the functions and to
substantiate claims made above.

\begin{figure}
\begin{center}
  \includegraphics*[width=0.49\textwidth]{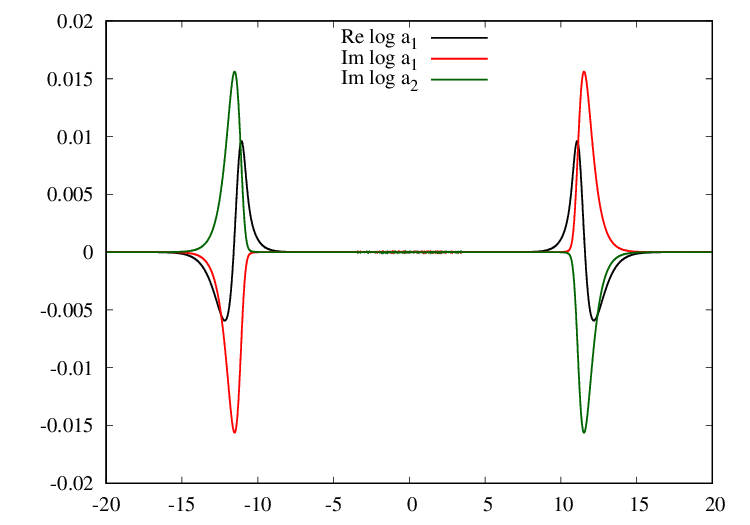}
  \includegraphics*[width=0.49\textwidth]{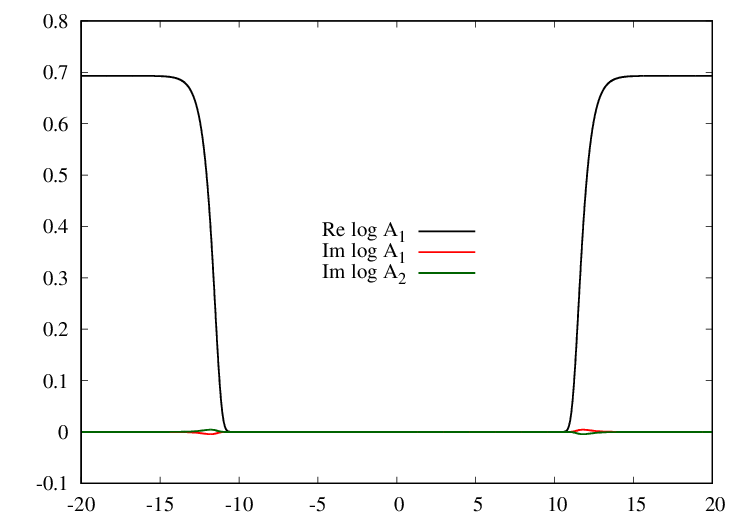}
  \caption{The true ground state of the system with $L/2$ roots located on the
    upper as well as on the lower axis: a) Depiction of real and imaginary
    parts of the functions $\log a_j-d$ with $j=1, 2, 3, 4$. The real parts
    are identical for all $j$,
    the imaginary parts are identical for
    $j=1, 3$ ($j=2, 4$).
    Similar depiction of the functions $\log A_j=\log(1+a_j)$. We use parameters $L=10^{10}$
  and $\gamma=0.8$.}
\label{Figloga0}  
\end{center}
\end{figure}

\begin{figure}
\begin{center}
  \includegraphics*[width=0.49\textwidth]{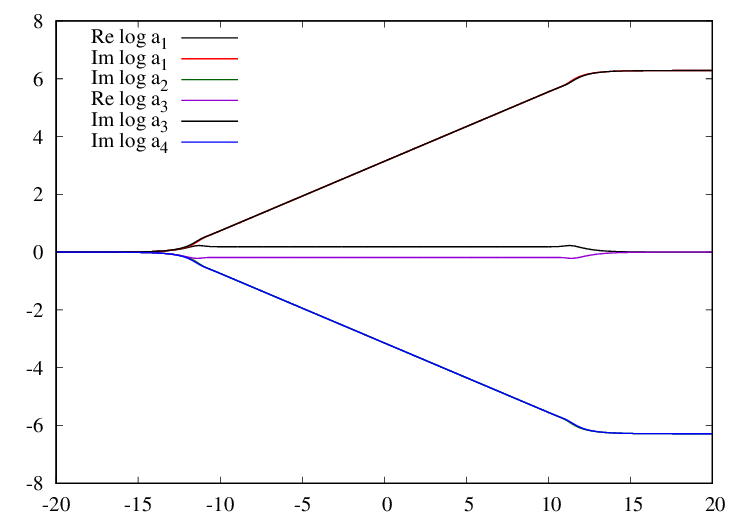}
  \includegraphics*[width=0.49\textwidth]{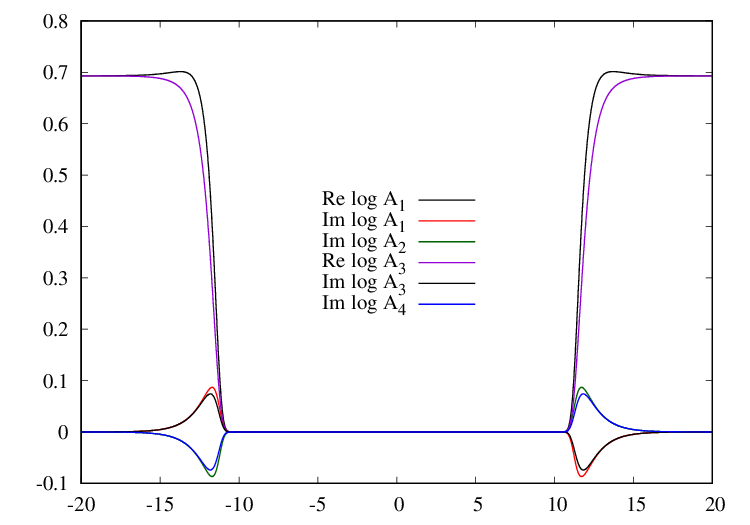}
 \caption{The first excited state ($n=1$) of the system with $L/2+1$ roots
   located on the upper axis and $L/2-1$ roots on the lower axis: a) Depiction
   of real and imaginary parts of the functions $\log a_i-d$ with $i=1, 2, 3,
   4$. The real parts for $i=1, 2$ ($i=3, 4$) are identical and shown by a
   black (violet) line. The imaginary parts for $i=1$ and $3$ ($i=2$ and $4$)
   are similar and shown by red and black (green and blue) lines. b) Similar
   depiction of the functions $\log A_i=\log(1+a_i)$.}
\label{Figloga1}
\end{center}
\end{figure}

In Fig.~\ref{Figloga0} we show graphs of the real and imaginary parts of the functions 
$\log a_i-d$ and $\log A_i=\log(1+a_i)$ for the true ground state solution and
system size $L=10^{10}$, $\gamma=0.8$.
Reallocating one root from one axis to the other ($n=\pm 1$) has a solution
that is shown in Fig.~\ref{Figloga1}.
Note that the changes in comparison to the ground state are minor for the
functions $\log A_i$, but there are huge changes in the $\log a_i$
functions. These functions no longer show equal asymptotics for argument to
$-\infty$ and $+\infty$. Instead they show a winding of the trajectories of the functions $a_i(x)$
around $0$: the asymptotics of $\Im \log a_i(x)$ change by $2\pi\ir$ from
$-\infty$ and $+\infty$ with almost linear dependence on the argument $x$ in the
interval $[-(\log L)/g,+(\log L)/g]$, see also the reasoning following (\ref{asymp_d}).

\section{Non-linear integral equations: regular kernel version}\label{sect:NLIEreg}

There are many alternative ways to write equation (\ref{singNLIE}), i.e.~with
not only $\log A_i$ on the right hand side, but also with $\log a_i-d$ functions. We found
that the following one is valid and useful
\begin{equation}
  \av=d+K_r\ast (\av-d-2\Av)
  \qquad \hbox{with}\quad K_r:=\frac K{K-2}. \label{regNLIE}
\end{equation}
The equivalence to (\ref{singNLIE}) is shown by straightforward ``algebraic'' rewriting.
The kernel $K_r$ has a block structure similar to $K$
\begin{equation}
\begin{split}
    K_r&=\left(\begin{matrix}
\kappa_1  &\kappa_2\cr
\kappa_2^\dagger & \kappa_1^T
\end{matrix}\right)\,, \qquad (\dagger\ \hbox{interchanges diagonal elements}).
\end{split}
\end{equation}
but unlike $K$ the entries are regular functions at $k=0$
\begin{equation}
\kappa_1(k)=
\frac{\sinh((\pi-2\gamma)k)}{2\sinh(\pi k)}
\left(\begin{matrix}
1  &-\er^{(\pi-2\gamma)k}\cr
-\er^{(2\gamma-\pi)k}& 1
\end{matrix}\right)\,, \qquad \kappa_2(k)= \frac{\sinh(2\gamma k)}{2\sinh(\pi k)}
\left(\begin{matrix} \er^{(\pi-2\gamma)k} & -1\cr -1& \er^{(2\gamma-\pi)k} \end{matrix}\right).
\end{equation}
The inverse Fourier transform yields (with a slight abuse of notation:
$\kappa_j(k)$ is the Fourier transform of $\kappa_j(x)$)
\begin{equation}
\kappa_1(x)=
\left(\begin{matrix}
\sigma(x)  &-\sigma(x-(\pi-2\gamma)\ir)\cr
-\sigma(x+(\pi-2\gamma)\ir)& \sigma(x)
\end{matrix}\right)\,, \qquad \kappa_2(x)= 
\left(\begin{matrix} -\sigma(x+ 2\gamma\ir+\ir\epsilon) & \sigma(x-\pi\ir)\cr
    \sigma(x+\pi\ir)&-\sigma(x- 2\gamma\ir-\ir\epsilon)
     \end{matrix}\right)\,,
\end{equation}
and
\begin{equation}
  \sigma(x):=\frac\ir2\left(\frac1{1-\er^{-x-2\gamma\ir}}-\frac1{1-\er^{-x+2\gamma\ir}}\right)
  =\frac{\sin2\gamma}{4\sinh(x/2+\gamma\ir)\sinh(x/2-\gamma\ir)}\,.
\end{equation}  
Generally, equation (\ref{regNLIE}) is of a type that allows for a direct
iterative method (like in~\cite{Klumper91,Damerau06}) to obtain, for arbitrary
lattice size $L$ and anisotropy $\gamma$, a numerical solution of high
accuracy. Among other things, for doing efficient computations of the
convolution integrals we use fast Fourier transform (FFT). The first question
that arises when looking at (\ref{regNLIE}) is how it is possible that all
singularities have disappeared and no problems -- possibly in different
clothing -- appear? The convolution integrals are indeed well-defined. The
right hand side of (\ref{regNLIE}) viewed as an operator has most eigenvalues
of absolute value below 1, except one eigenvalue at $k=0$ which is exactly
$+1$. The corresponding eigenvector, however, does not realize an instability
to convergence in the iterative numerical calculations. This is largely
prohibited by the bulk behaviour of the functions in the interval $[-(\log
  L)/g,+(\log L)/g]$. Next to mention is the peculiar factor 2 in the
combination $a-d-2A$. By this, the functions $a-2A$, or in extended notation
$\log a_i(x)-2\log A_i(x)$, have very quickly decaying asymptotics for $x$ to
$\pm\infty$. And last, the combination $a-d$ has numerically small values as
$d$ is the bulk behaviour of $a$.

Here a note on the relation of our integral equations to those used by
\cite{Candu_2013,Frahm14} is in order. In \cite{Candu_2013} the NLIE of the
form (\ref{singNLIE}) is used, the leading singular behaviour of the kernel is
extracted resulting in additional source terms. The so treated NLIE in the
scaling limit is then solved numerically. The work \cite{Frahm14}
uses a different version of NLIE of course related to (\ref{singNLIE})
resp.~\cite{Candu_2013}. The NLIEs of \cite{Frahm14} can be obtained 
with functions on contours similar to (\ref{deflittlea1})-(\ref{deflittlea4})
however without taking the reciprocal of the function $a$ in the definition of
$a_1$ and $a_3$. Following the steps presented above we arrive at NLIEs of the
form (\ref{singNLIE}) for the modified functions $\tilde a_j$ and $\tilde
A_j=1+\tilde a_j$ with a different kernel which appears to be
regular. Technically, in \cite{Frahm14} the NLIEs are presented in closed
contour formulation. 

We find that the form (\ref{regNLIE}) is highly useful for treating
unprecedented ranges of values for $L$ and in principle any lattice size we
want with high numerical accuracy. In this paper we limit our numerics to
values of $L$ between 2 and $10^{24}$. For efficient numerics we use the Fast
Fourier Transform for calculating the convolution integrals. To this end the
eqns.~(\ref{regNLIE}) need the application of certain subtraction terms to
render the functions entering the convolutions Fourier transformable. This
will be treated in the next paragraph. A final remark on the two different
versions of the NLIEs that we deal with: the ``singular'' version of the NLIE
(\ref{singNLIE}) is not at all useless. It is important to note that the
singular nature of the kernel is closely connected to the emergence of a
continuous spectrum in the continuum limit of the model as discussed by Candu
and Ikhlef in~\cite{Candu_2013}. It will allow for an analytic derivation of
at least the leading logarithmic terms in the large size asymptotics, see
Sect.~\ref{Sect:AnalyticalStudy}.

Note that (\ref{regNLIE}) like (\ref{singNLIE}) holds for any imbalance $n$ of
the number of roots on the axes with imaginary parts $\pm\pi/2$. For $n=0$ the
solution functions have simple asymptotics as shown in Fig.~\ref{Figloga0}.
However, shifting $n$ BA roots from one axis to the other one yields a winding
of the $\log a_i(x)$ functions: $\log a_i(\infty)-\log a_i(-\infty)=\pm n\,
2\pi\ir$. Interestingly, (\ref{regNLIE}) holds for any $n$, but does not contain
any $n$ dependent term. The convolution integral in (\ref{regNLIE}) is
well-defined for any $n$, but does not allow for an application of  the
Fourier transform, because of the different asymptotes for non-zero $n$. This
problem is cured by subtracting and adding analytic functions $w$ and $\tilde
w$ with the property $w=K_r\ast \tilde w$, such that
\begin{align}
  a&=d+K_r\ast (a-d-2A)
=d+n\,w +K_r\ast (a-d-n\,\tilde w-2A),\label{finalNLIE}
\end{align}
and $a-n\,\tilde w$ has same asymptotical values at $-\infty$
and at $+\infty$.
This is achieved by for instance the functions
\begin{equation}
  w(x)=
  \left(\begin{matrix}
    w_1(x)\\
    w_2(x)\\
    w_3(x)\\
    w_4(x)
  \end{matrix}\right),\quad  
  \tilde w(x)=2\log\tanh\left(\frac{g}{2} x+\ir\frac\pi4\right)
\cdot  \left(\begin{matrix}
    +1\\
    -1\\
    +1\\
    -1
  \end{matrix}\right),
\end{equation}
and
\begin{align}
w_1(x)&=-w_4(x):=\log\tanh\tfrac12\left(x+\ir\left(\frac\pi2-\gamma\right)\right)
+\log\tanh\tfrac12\left(x+\ir\left(3\gamma-\frac\pi2\right)\right),\\
w_2(x)&=-w_3(x):=\log\tanh\tfrac12\left(x-\ir\left(\frac\pi2-\gamma\right)\right)
+\log\tanh\tfrac12\left(x-\ir\left(3\gamma-\frac\pi2\right)\right).
\end{align}
In our numerical calculations we have used variations of these functions like
convolutions of the $w_j$, $\tilde w_j$ with functions that increase linearly
in the interval $[-(\log L)/g,+(\log L)/g]$ such that the modified $w_j$,
$\tilde w_j$ are still smooth, but mimic the behaviour of the functions
$\log\av_j$. This leads to a slight improvement in the numerical accuracy.

Eq.~(\ref{finalNLIE}) is the final version of NLIE that we use for the numerical
treatment of the low-lying energy states as long as the distributions of the
zeros of $q$, i.e.~the Bethe roots, and of the zeros of the eigenvalue
function $\Lambda$ are as we described above.

We now use the number $n$, i.e.~the winding number of the functions $a_i$, as
the fundamental variable. The quasi-momentum, see \cite{ikhlef2012conformal} and
(\ref{quasimom}), is a derived property and does not play a central role in our
treatment.

Within the presented treatment it is also possible to obtain a simple expression for the
energy in terms of the auxiliary functions $\log A_i$. For details see the Appendix.
Here we give the integral expression for the relevant combination of the
eigenvalue functions which splits into a pure bulk and a finite size part
\begin{equation}
\log[\Lambda(x-\ir\gamma)\Lambda(x+\ir(\pi-\gamma))]=
  L\cdot \lambda_0(x)+\kappa\ast[\log \Av_1+\log \Av_2+\log \Av_3+\log \Av_4].\label{LamEnerg}
\end{equation}
The kernel function $\kappa$ is given by
\begin{equation}
    \kappa(x)=-\ir\frac g{\sinh(g x)}.
\end{equation}
The energy is calculated from the derivative at $x=0$
\begin{align}
   E&=\sin(2\gamma)\i\frac{d}{dx}\log[\Lambda(x-\ir\gamma)\Lambda(x+\ir(\pi-\gamma))]\\
&=L e_0-\frac{\sin(2\gamma)}{2\pi} \int_{-\infty}^\infty dx \,\frac{g^2 \cosh gx}{(\sinh gx)^2}\,[\log \Av_1(x)+\log \Av_2(x)+\log \Av_3(x)+\log \Av_4(x)] ,\label{energyintegral}
\end{align}
where $e_0$ is the bulk energy obtained from $\lambda_0(x)$, and we remind of
the definition of $g$ in (\ref{abbreviations}).

The so-called quasi-momentum $K$ \cite{ikhlef2012conformal} is the logarithm of the quasi-shift operator
\begin{equation}
\tilde\tau:=T(\pi/2)T^{-1}(0)\,.\label{quasimom}
\end{equation}
The eigenvalue for a given state is obtained from the function
\begin{equation}
K(x):=\log\frac{\Lambda(x+\ir(\pi-\gamma))}{\Lambda(x-\ir\gamma)}=
\tilde\kappa\ast[\log \Av_1-\log \Av_2+\log \Av_3-\log \Av_4],\qquad
\tilde\kappa(x)=\ir g \,{\coth(g x)},\label{defK}
\end{equation}
at argument $x=0$
\begin{equation}
K=K(0)=\frac g{2\pi\ir}\int_{-\infty}^\infty dx \,\coth gx\,[\log \Av_1(x)-\log \Av_2(x)+\log
  \Av_3(x)-\log \Av_4(x)].\label{qmomentum}
\end{equation}

As the model is critical the spectrum of the low-lying excitations can be
described within the framework of CFT. The finite size parts were calculated
previously by different approaches~\cite{ikhlef2012conformal,Candu_2013,Frahm14,bazhanov2019scaling}
\begin{equation}
E(L)=L\,e_0+\frac{2\pi}L
v_F\left(-\frac16+\frac\gamma{2\pi}m^2+\frac\pi{2\gamma}w^2
+\frac{2g\gamma}{\pi}s^2\ +\ N\right),
\qquad v_F=g\sin(2\gamma),\label{extCFT}
\end{equation}
which means that the (effective) central charge is $c=2$. Here $v_F$ is the Fermi velocity
and $m$ and $w$ are integers identical to the magnetization (number of flipped
spins) and to the imbalance of the root distributions between left and
right, $N$ is the integer labeling the levels of the conformal tower. The case
we studied above corresponds to $m=w=N=0$.

The energy contributions due to the parameter $s$ originate from rather
remarkable logarithmic finite-size corrections as a consequence of the
reallocation of the number $n$ of BA-roots from one axis to the other. The
Wiener-Hopf analysis in \cite{ikhlef2012conformal} resulted in
\begin{equation}
 s\simeq\frac{\pi\, n}{2\log L} \quad \hbox{for large}\ L,\quad \ n=0, \pm 1,\label{asymp_for_s}
 \pm 2,...
\end{equation}
In the thermodynamical limit this term signifies a non-compact degree of
freedom as the dicretization of $s$ is on the logarithmic scale
${1}/{\log(L)}$ and to be considered a continuous parameter.

The result for the quasi-momentum in terms of $s$ is
\begin{equation}
K={4g\gamma}\, s.\label{quasmoms}
\end{equation}
To illustrate, we obtain the result (\ref{extCFT},\ref{asymp_for_s}) for the energy and the
quasi-momentum in Sect.~\ref{Sect:AnalyticalStudy} by
simple manipulations starting from (\ref{singNLIE}) without solving any NLIEs,
not to mention using Wiener-Hopf techniques.

\begin{figure}
\begin{center}
  \includegraphics[scale=0.5]{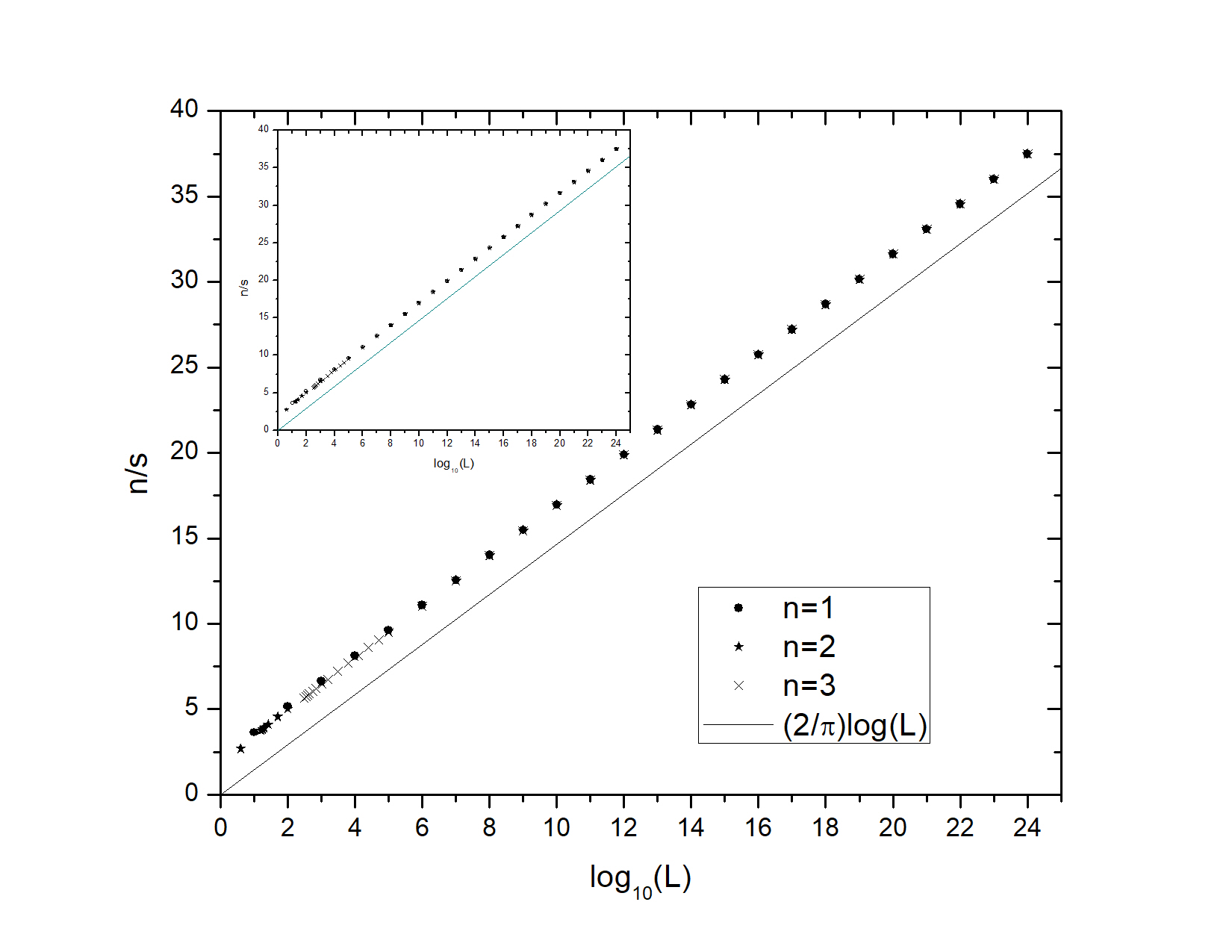}
  \caption{Plot of the ratio $n/s$ against system size $L$ with quasi-momentum
    parameter $s$ obtained from calculations of energies $E_{NLIE}$ (main
    panel) and quasi-momenta $K_{NLIE}$ (inset) by use of the NLIE
    (\ref{finalNLIE}). The continuous line corresponds to the asymptotic
    behavior (\ref{asymp_for_s}). The symbols lie above this line by an offset
    $\log L_0$ defining the non-universal length scale $L_0$. The data have
    been obtained for parameters $\gamma=0.8$ and $n=1,2,3$.}
\label{fig:plot_s}
\end{center}
\end{figure}

In Fig.~\ref{fig:plot_s} we show the results of our calculations for various
values of $n$ and system sizes from $L=2$ to $L=10^{24}$. The data obtained
for the energy resp.~the quasi-momentum are used to identify the parameter $s$
from (\ref{extCFT}) resp.~(\ref{quasmoms}). We plot $n/s$ as function of $L$
and see clearly the asymptotic behaviour (\ref{asymp_for_s}).

The finite, i.e.~next order term to (\ref{asymp_for_s}) resp.~to the density
of states ($\partial n/\partial s$) was obtained in \cite{Candu_2013,Frahm14}
by numerical calculations based on the NLIEs used by the authors.  The most
accurate analytical results for the asymptotics of $s$, in the form of a
quantization condition, are given in \cite{ikhlef2012conformal} and
especially in \cite{bazhanov2019scaling} by use of the ODE/IQFT
correspondence.

\begin{figure}
\begin{center}
  \includegraphics[scale=0.5]{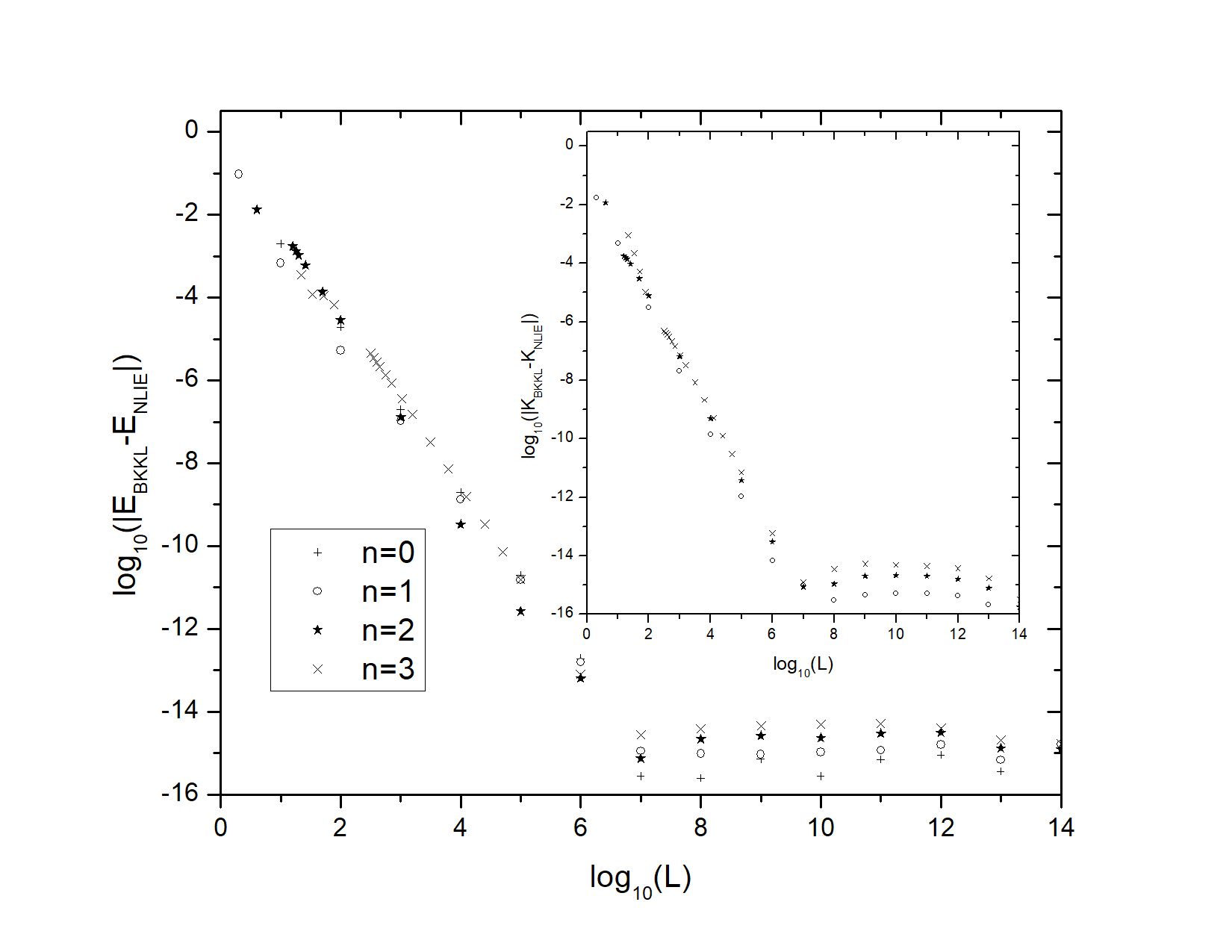}
\caption{Comparison of the energies $E_{NLIE}$ and quasi-momenta $K_{NLIE}$
  (inset) from numerically solving the NLIE (\ref{finalNLIE}) with
  $E_{BKKL}=({2g\gamma}/{\pi})\,s^2$ and $K_{BKKL}=4g\gamma\,s$ (inset) with
  $s$ computed from the quantization condition in
  \cite{bazhanov2019scaling}. Plotted is the difference which vanishes algebraically
  with system size like $\mathcal{O}(L^{-2})$, but not below values of the order
  $10^{-15}$ to $10^{-14}$ set by the accuracy of our numerical calculations
  with double precision operations. The parameters were taken to be
  $\gamma=0.8$ and $n=0,1,2,3$.}
\label{fig:Energy}
\end{center}
\end{figure}

In Fig.~\ref{fig:Energy}, we compare energy and quasi-momentum for different
values of $n$ and various system sizes obtained from a numerical solution of
(\ref{regNLIE}) with the results of \cite{bazhanov2019scaling}, see also
(\ref{ODEIQFT1})-(\ref{ODEIQFT4}) in the appendix.  The values of the parameters
used in the plot are $\gamma=0.8$ and $n=0,1,2,3$. Our NLIE are exact for all
(even) lattice sizes even for the smallest value $L=2$.  The quantization
condition \cite{bazhanov2019scaling} has been derived for large sizes, but
interestingly it is rather accurate also for small lattice sizes. The
difference of the two approaches is best fitted by an order
$\mathcal{O}(L^{-2})$ ansatz (for $\pi/4<\gamma<\pi/2$ where we carried out
our numerical calculations).

\section{ Analytical study of the scaling dimensions}\label{Sect:AnalyticalStudy}
To handle the scaling limit for $L \rightarrow \infty$, we observe that in the
NLIEs (\ref{singNLIE},\ref{regNLIE}) only the driving term explicitly depends on $L$ and
possesses the asymptotics
\begin{equation}
   {L} \log\tanh\frac{g}{2}x\simeq -2L\er^{-g|x|},\label{asymp_d}
\end{equation}
from which we conclude that all functions $a_j(x)$ are extremely small for
$|x|<(\log L)/g$ and change to their asymptotic form in the neighbourhood of
$x\sim\pm(\log L)/g$. This property may be used to define non-trivial
functions in the scaling limit by suitable shifts of the argument by the
offset $\pm(\log L)/g$. Here, due to the necessity to work with the functions
on the positive and negative real semi-axes at the same time, we avoid these
definitions.

We next use the differentiated NLIE with singular kernel
\begin{equation}
(\log \av_i)'=d'+\sum_{j=1}^4K'_{ij}\ast\log(1+\av_j),
\end{equation}
which we multiply from left by $\log(\Av_i(x))$, sum over the index $i$ and
integrate along the positive semi-axis
\begin{align}
  \int_0^\infty dx\sum_{i=1}^4\log(\Av_i(x))(\log \av_i(x))'=&
  \int_0^\infty dx\sum_{i=1}^4\log(\Av_i(x))d'(x)\cr&+
\frac1{2\pi}  \int_0^\infty dx  \int_{-\infty}^\infty dy\sum_{i,j=1}^4
\log(\Av_i(x))K'_{ij}(x-y)\log(\Av_j(y)).\label{doubleintegr}
\end{align}
The next steps consist in showing: (i) the integral on the left hand side can be
explicitly evaluated, (ii) the first term on the right hand side is -- up to a scale
factor -- identical to the energy integral in (\ref{energyintegral}), (iii)
the second term on the right hand side can also be evaluated. These calculations are done
for large system size $L$ and ignore certain higher order terms.

Ad (i): a change of the variable of integration from $x$ to $a_i$ turns the
left hand side of (\ref{doubleintegr}) into a dilogarithmic integral along the trajectory
the function $a_i(x)$ takes from $x=0$ to $x=+\infty$, namely from $a_i(0)=0$ to
$a_i(+\infty)=1$. The trajectory itself does not matter as long as it does not
wind around the singularities of the integrand, which we asume here. For each $i$
the integral evaluates to $\pi^2/12$, times 4 yielding $\pi^2/3$. The error done
is of the order $a_i(0)$, that is exponentially small in $L$.

Ad(ii):
The first term on the right hand side is, with a view to (\ref{asymp_d}) equal to
\begin{equation}
L2g \int_{0}^\infty dx\, \er^{-gx}\,[\log
  \Av_1(x)+\log \Av_2(x)+\log \Av_3(x)+\log \Av_4(x)],
\end{equation}
with an error of order $\mathcal{O}(L^{-2})$.
The integral in (\ref{energyintegral}) along the positive semi-axis is
identical with a different prefactor
\begin{equation}
-\sin(2\gamma)\frac{g^2}\pi \int_{0}^\infty dx\, \er^{-gx}\,[\log
  \Av_1(x)+\log \Av_2(x)+\log \Av_3(x)+\log \Av_4(x)],
\end{equation}
and an error of order $\mathcal{O}(L^{-3})$.

Ad (iii):
Here we use the fact that the kernel matrix $K_{ij}(x-y)$ is
symmetric with respect to an exchange of $x,i$ and $y,j$, and the
differentiated kernel $K'_{ij}(x-y)$ is antisymmetric. The second term on
the right hand side of (\ref{doubleintegr}) can be massaged
\begin{equation}
  \int_0^\infty dx  \int_{-\infty}^\infty dy\sum_{i,j=1}^4...=
  \underbrace{\int_0^\infty dx  \int_{0}^\infty dy\sum_{i,j=1}^4...}_{=0}
  +  \int_0^\infty dx  \int_{-\infty}^0 dy\sum_{i,j=1}^4...\label{doubleintsplit}
\end{equation}
where $...$ refer to the integrand in (\ref{doubleintegr}) with antisymmetry
in $x,i$ and $y,j$ and thus the first term with symmetric integrations and
sums over $x,i$ and $y,j$ yields zero.

In the second term on the right hand side of (\ref{doubleintsplit}) the variable of
integration is a positive $x$ (negative $y$). The integrand is only
noticeably different from zero for $x$ ($y$) taking values around $(\log L)/g$
($(-\log L)/g$) and larger (lower). This means that $K_{ij}(x-y)$ can be
replaced by its asymptotic behaviour as it matters only for $x-y$ taking
values around $2(\log L)/g$ or larger.
Replacing $K'_{ij}$ by the limiting values $(-1)^{i+j}g/2\gamma$ yields
\begin{align}
\int_0^\infty dx \int_{-\infty}^0 dy\sum_{i,j=1}^4
\log(\Av_i(x))K'_{ij}(x-y)\log(\Av_j(y))&=\frac{g}{2\gamma}\int_0^\infty
dx \int_{-\infty}^0 dy\sum_{i,j=1}^4(-1)^{i+j}
\log(\Av_i(x))\log(\Av_j(y))\nonumber\\
&=\frac{g}{2\gamma}|I|^2,\label{squareI}
\end{align}
where we define (the identity is shown below):
\begin{equation}
  I:=\int_0^\infty dx\log\left(\frac{\Av_1(x)\Av_3(x)}{\Av_2(x)\Av_4(x)}\right)
  =-\int_{-\infty}^0
  dx\log\left(\frac{\Av_1(x)\Av_3(x)}{\Av_2(x)\Av_4(x)}\right).
  \label{defI}
\end{equation}
In this way the problem reduces to calculate the integral $I$, which can be
obtained from the NLIE
\begin{align}
   2\pi\i n &=\log \av_1(+\infty)-\log
   \av_1(-\infty)\\ &=\frac1{2\pi}\lim_{x \rightarrow\infty}
   \int_{-\infty}^\infty dy\sum_{j=1}^4
   K_{1j}(x-y)\log(\Av_j(y))-\frac1{2\pi}\lim_{x
     \rightarrow-\infty} \int_{-\infty}^\infty dy\sum_{j=1}^4
   K_{1j}(x-y)\log(\Av_j(y))\\ &=\frac g{4\gamma\pi}\left[\lim_{x
       \rightarrow\infty} \int_{-\infty}^\infty dy\,
     (x-y)\log\left(\frac{\Av_1(y)\Av_3(y)}{\Av_2(y)\Av_4(y)}\right)+\lim_{x
       \rightarrow-\infty} \int_{-\infty}^\infty dy\,
   (x-y)\log\left(\frac{\Av_1(y)\Av_3(y)}{\Av_2(y)\Av_4(y)}\right)\right].
\end{align}
From this resp.~the existence of the $x\to\pm\infty$ limits we learn that the integral over
$\log[\Av_1(y)\Av_3(y)/\Av_2(y)\Av_4(y)]$ from $y=-\infty$ to $+\infty$ is
zero and drops out. This also proves the identity in (\ref{defI}).
The remaining integrals give
\begin{equation}
2\pi\i n=...=-\frac g{2\gamma\pi}\int_{-\infty}^\infty dy\,
   y\log\left(\frac{\Av_1(y)\Av_3(y)}{\Av_2(y)\Av_4(y)}\right)
=-\frac{g}{\gamma\pi}\frac{\log
       L}g\cdot I,
\end{equation}
where we have used in the last step the fact that
$\log[\Av_1(y)\Av_3(y)/\Av_2(y)\Av_4(y)]$ is purely
imaginary and non-zero only in the neighborhood of $y=(\log L)/g$.
From the last equation we find
\begin{equation}
  I=-\ir\frac{2\pi^2\gamma}{\log L}n,\label{Ileading}
\end{equation}
and hence the double integral (\ref{doubleintsplit}) is
\begin{equation}
    \int_0^\infty dx  \int_{-\infty}^\infty dy\sum_{i,j=1}^4
    \log(\Av_i(x))K'_{ij}(x-y)\log(\Av_j(y)) =
    {2g\gamma}\left(\frac{\pi^2 n}{\log L}\right)^2
\end{equation}
The leading errors in the calculation of (\ref{squareI}) are of order
$\mathcal{O}\left(L^{-(\pi-2\gamma)/\gamma}\right)$ and $\mathcal{O}\left(L^{-1}\right)$. There is
however a special structure of these terms and it seems likely that they
recombine and leave (\ref{squareI}) correct to high order in powers of
$L^{-1}$ with additional terms to $I$ (\ref{defI}). The calculation of $I$ in
(\ref{Ileading}) comes of course with leading errors resulting in changing
$\log L$ to a $\log L/L_0$ with some constant $L_0$.  Summing the above terms
(i), (ii), (iii) with the analogues resulting from (\ref{doubleintegr}) with
integration over $x$ along the negative semi-axis we obtain
\begin{equation}
\frac23\pi^2=-\frac{2\pi L}{g\sin(2\gamma)}(E-L
e_0)+\frac{2g\gamma}{\pi}\left(\frac{\pi^2n}{\log L}\right)^2,
\end{equation}
which is (\ref{extCFT}) for $m=w=N=0$ with (\ref{asymp_for_s}).

\section{Deformed patterns of root distributions / Descendant states}\label{sect:Desc}
In this section we study states with root distributions that show qualitative
differences in comparison to those shown in
Fig.~\ref{fig:zeros-q-Lambda}. Here we understand by root distribution the
distribution of zeros of $q(z)$ and those of $\Lambda(z)$.
For such roots not lying in the ``standard'' regions of the complex plane
particular care has to be applied leading to NLIE with additional terms in the
driving term, see e.g.~\cite{Klu93,ahn2005nlie}.

\begin{figure}
\begin{center}
\includegraphics[scale=0.3]{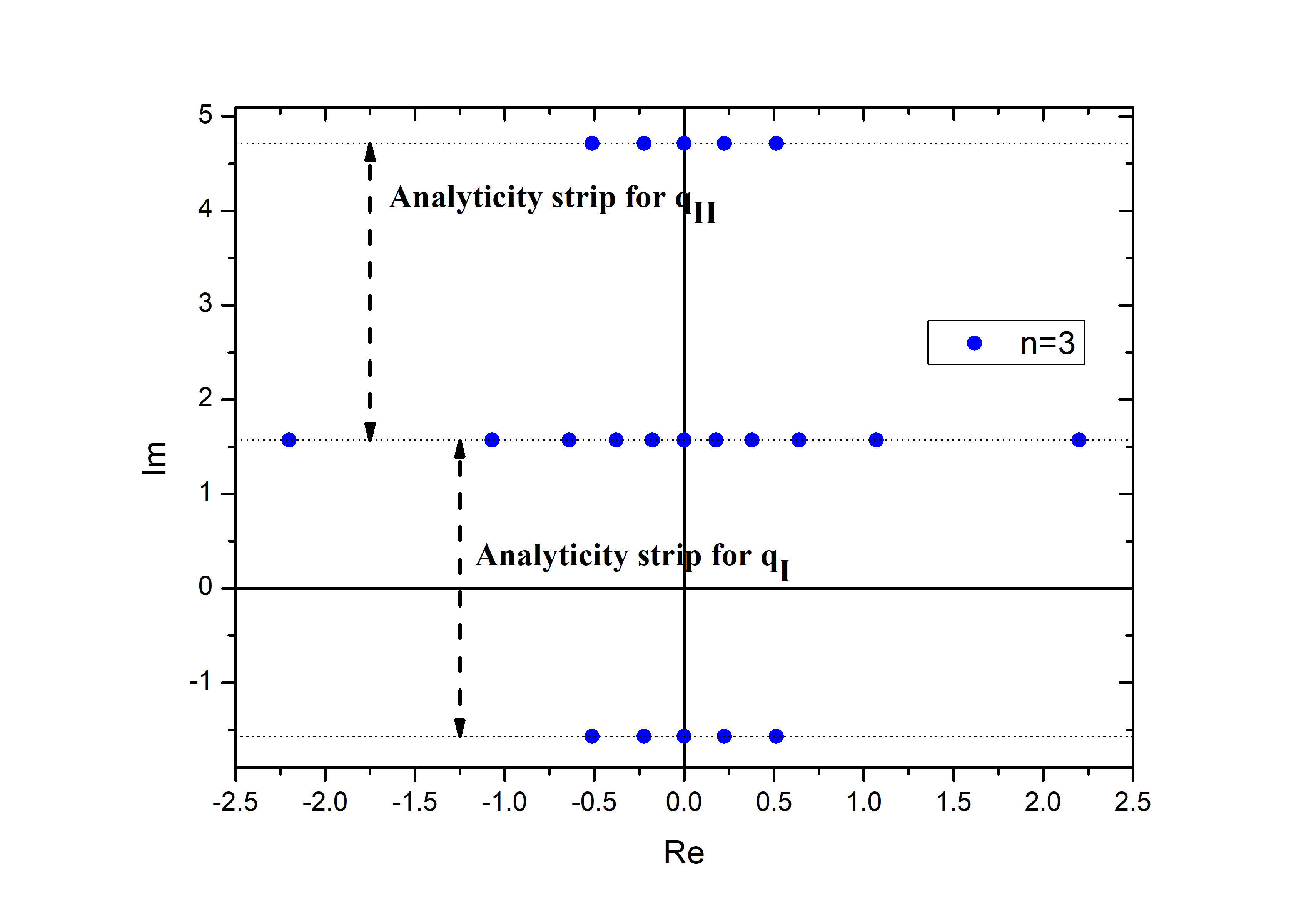}
\includegraphics[scale=0.3]{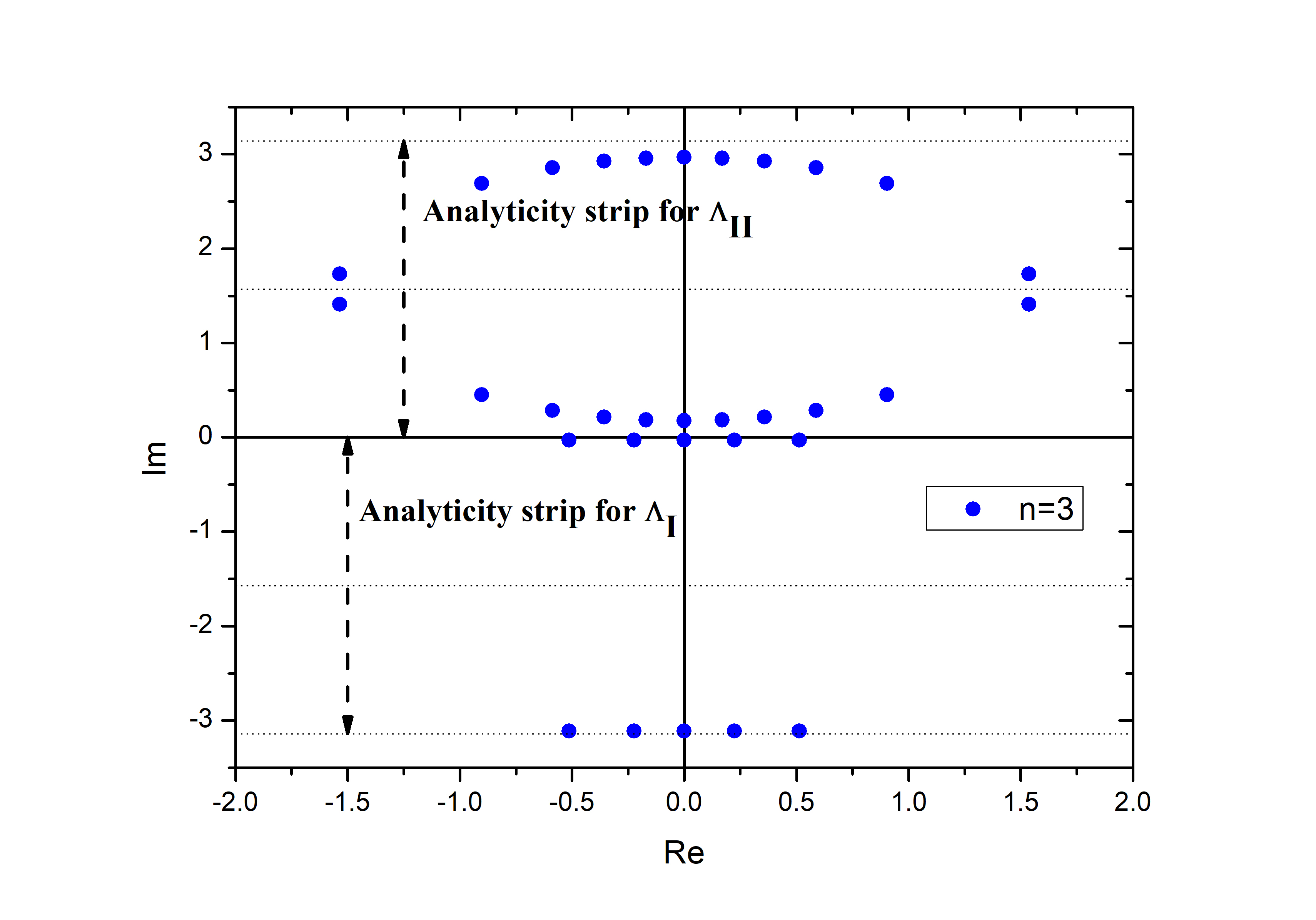}
\caption{Depiction of the location of all zeros like in  Fig.~\ref{fig:zeros-q-Lambda}
for parameters $L=16$ and $\gamma=0.8$. Here the value of $n=3$ is large against
  $\log L$ resulting in strong deviations of the zeros of $\Lambda(z)$ from the lines
$\mathbb{R}$ and $\mathbb{R}\pm \ir{\pi}$. Four of these zeros deviate
extremely and have to be treated explicitly as described in the main text.
Note that the zeros of $q(z)$ still lie precisely on straight lines.}
\label{fig:zeros-q-Lambda_large}
\end{center}
\end{figure}

The first such case, interestingly, corresponds to the class of states treated
in Sect.~\ref{sect:NLIEsing} resp.~Fig.~\ref{fig:zeros-q-Lambda}, however with
a large ratio $n/\log L$. In this case, the zeros of $q(z)$ are still
exclusively distributed along the lines $\mathbb{R}\pm \ir{\pi}/{2}$, but some
zeros of $\Lambda(z)$ deviate strongly from the lines $\mathbb{R}$ and
$\mathbb{R}\pm \ir{\pi}$, see ~Fig.~\ref{fig:zeros-q-Lambda_large}, such that
the Fourier transforms of the logarithmic derivatives of the $\Lambda$
functions appearing in (\ref{defbigA1})-(\ref{defbigA4}) can no longer be
reduced to just two independent functions.

Here we treat the case that four zeros, called $\thet_1, ..., \thet_4$, enter
the region of the complex plane with imaginary parts between $\gamma$ and
$\pi-\gamma$. In (\ref{defbigA1}) and (\ref{defbigA2}) the function
$\Lambda(z)$ is evaluated on the lines $\mathbb{R}+\ir\gamma$ and
$\mathbb{R}+\ir(\pi-\gamma)$ with four zeros inbetween. Therefore the Fourier
transform of the logarithmic derivatives of these equations does not lead to
(\ref{FTA1}) and (\ref{FTA2}) in the appendix. Of course, (\ref{FTA3}) and (\ref{FTA4}) are
still valid. Analogous considerations apply for zeros with imaginary parts
between $-(\pi-\gamma)$ and $-\gamma$.

There are several ways to proceed in this situation. Here we use
subtraction terms. We define
\begin{equation}
\lambda(z):=\prod_{j=1}^4\sinh\tfrac12(z-\thet_j),\qquad \hbox{(version 1)}\label{counterver1}
\end{equation}  
as well as modified functions 
\begin{equation}
  \tilde A_1(x):=\frac{A_1(x)}{\lambda(x+\ir\gamma)},
  \quad\tilde A_2(x):=\frac{A_2(x)}{\lambda(x+\ir\pi-\ir\gamma)},
  \quad\tilde A_3(x):=\frac{A_3(x)}{\lambda(x-\ir\gamma)},
  \quad\tilde A_4(x):=\frac{A_4(x)}{\lambda(x+\ir\pi+\ir\gamma)}.\label{tildeAfunctions}
\end{equation}  
These functions satisfy factorizations like (\ref{defbigA1})-(\ref{defbigA4}) with the
only change that $\Lambda(z)$ is replaced by
\begin{equation}
\tilde\Lambda(z):=\frac{\Lambda(z)}{\lambda(z)},
\end{equation}  
which has no zeros in the region discussed above. Hence the Fourier transforms
of all equations yield (\ref{FTA1})-(\ref{FTA4}) where now $\hToo$, $\hTot$
are the Fourier transforms of the logarithmic derivative of $\tilde\Lambda(z)$
in regions I and II. From this again (\ref{singNLIE}) and
(\ref{abbreviations}) follow with the only modification that $A_j$ are to be
replaced by $\tilde A_j$. The same holds for (\ref{regNLIE}) and (\ref{finalNLIE}).

A problem is left with (\ref{counterver1}), which renders the functions
(\ref{tildeAfunctions}) having vanishing asymptotics, so their logarithms have
diverging asymptotics. These functions do not allow for the Fourier transform
which is necessary for the numerics. Hence we modify the definition
(\ref{counterver1}) -- by keeping (\ref{tildeAfunctions}) -- to
\begin{equation}
\lambda(z):=\prod_{j=1,3}\frac{\sinh\tfrac12(z-\thet_j)}{\sinh\tfrac12(z-\Re\thet_j)}
\prod_{j=2,4}\frac{\sinh\tfrac12(z-\thet_j)}{\sinh\tfrac12(z-\Re\thet_j-\pi\ir)}
,\qquad \hbox{(version 2)},\label{counterver2}
\end{equation}  
where we assumed that $0<\Im\thet_{1,3}<\pi/2$ and $\pi/2<\Im\thet_{2,4}<\pi$.

Now we can use (\ref{finalNLIE}) with the replacement of $A_j$ by $\tilde
A_j$. Of course the location of all $\thet_j$ has to be known. They are
determined from the condition $a(\thet_j)=-1$ or for instance
from $a_1(\thet_j-\ir\gamma)=-1$ which is evaluated by use of the NLIEs as
integral expressions for the functions $\log a_j(x)$ allowing for complex
arguments $x$.

This program leads to a coupled set of NLIEs and scalar equations for the
$\thet_j$. Having solved it, the integral expressions for energy and
quasi-momentum have to be evaluated. To this end we may start with
(\ref{LamEnerg}) and (\ref{defK}) with $\Lambda$ and $\Av_j$ replaced by
$\tilde\Lambda$ and $\tilde\Av_j$, and then obtain the analogues of
(\ref{energyintegral}) and (\ref{qmomentum}) with additional $\thet_j$-terms
appearing on the right hand sides. The modified (\ref{energyintegral}) is
still an integral over the product of a function with pole of 2nd order in
$x=0$ times the sum over all $\tilde\Av_j$ which, unlike the sum over all
$\Av_j$ does not have a zero of high order at $x=0$. Therefore, the resulting
expressions need to be written in terms of integrals involving the $\Av_j$.

The obtained modified versions of
(\ref{energyintegral}) and (\ref{qmomentum}) are
\begin{align}
  E&=L e_0
+\sum_{j=1}^4\frac{g\sin(2\gamma)}{\cosh\left(g\left(\thet_j-\frac\pi2\ir\right)\right)}
  -\frac{\sin(2\gamma)}{2\pi} \int_{-\infty}^\infty dx \,\frac{g^2 \cosh
    gx}{(\sinh gx)^2}\sum_{j=1}^4\log \Av_j(x) ,\label{energyintegralmod}\\
  K&=\log\left[\prod_{j=1}^4\sinh(g(\thet_j-\ir\gamma))\right]+
  \frac g{2\pi\ir}\int_{-\infty}^\infty dx \,\coth(gx)\,
\log\left[\tanh^8\left(\frac g2x+\frac\pi4\ir\right)\cdot\frac{\Av_1(x)\Av_3(x)}{\Av_2(x)\Av_4(x)}\right],
\label{qmomentummod}
\end{align}
where here indeed the functions $\Av_j$ enter, not the $\tilde\Av_j$.  We find
that equation (\ref{energyintegralmod}) is obtained most elegantly from the
original (\ref{energyintegral}) which also holds in the currently considered
case provided the contours for the integrals are suitably deformed. In the
convolution integral an integration by parts is carried out resulting in a
convolution over the kernel function $\kappa$ and the derivative of the sum of
$\Av_j$. Then by use of Cauchy's theorem the deformations are ``straightened''
upon which contributions of residues appear as the new explicit terms in
(\ref{energyintegralmod}).  At last, in the convolution integral with straight contours
the integration by parts is run backwards to give (\ref{energyintegralmod}).
For the derivation of (\ref{qmomentummod}) we
start from (\ref{qmomentum}) with deformed contours and perform an integration
by parts where we use $\tilde\kappa(x)=\ir g \,{\coth(g x)}$
\begin{equation}
K=\frac{-1}{2\pi\ir}\int dx(\log\sinh gx)[(\log \Av_1(x))'-(\log \Av_2(x))'+(\log
  \Av_3(x))'-(\log \Av_4(x))'],\quad
\hbox{(deformed contours!)}.\label{defKdiff}
\end{equation}
We straighten the deformed contours and obtain
\begin{align}
  K&=\sum_{j=1}^4\log\sinh(g(\thet_j-\ir\gamma))\cr
  &-\frac1{2\pi\ir}\int dx(\log\sinh gx)[(\log \Av_1(x))'-(\log \Av_2(x))'+(\log
  \Av_3(x))'-(\log \Av_4(x))'],\label{defKdiff2}
\end{align}
with straight contours. It is tempting to perform directly an integration by
parts.  However, the resulting integral does not exist, because the function
$\log \Av_1-\log \Av_2+\log \Av_3-\log \Av_4$ has non trivial
asymptotics. Adding and subtracting the $\log\tanh$ term appearing in
(\ref{qmomentummod}) cures the problem: the integral with just the counter
term can be done and for argument $x=0$ yields zero! The remaining integral with the modified
second factor allows for integrating by parts. This way we obtain
(\ref{qmomentummod}) with well-defined integral.\\

\begin{figure}
\begin{center}
\includegraphics[scale=0.3]{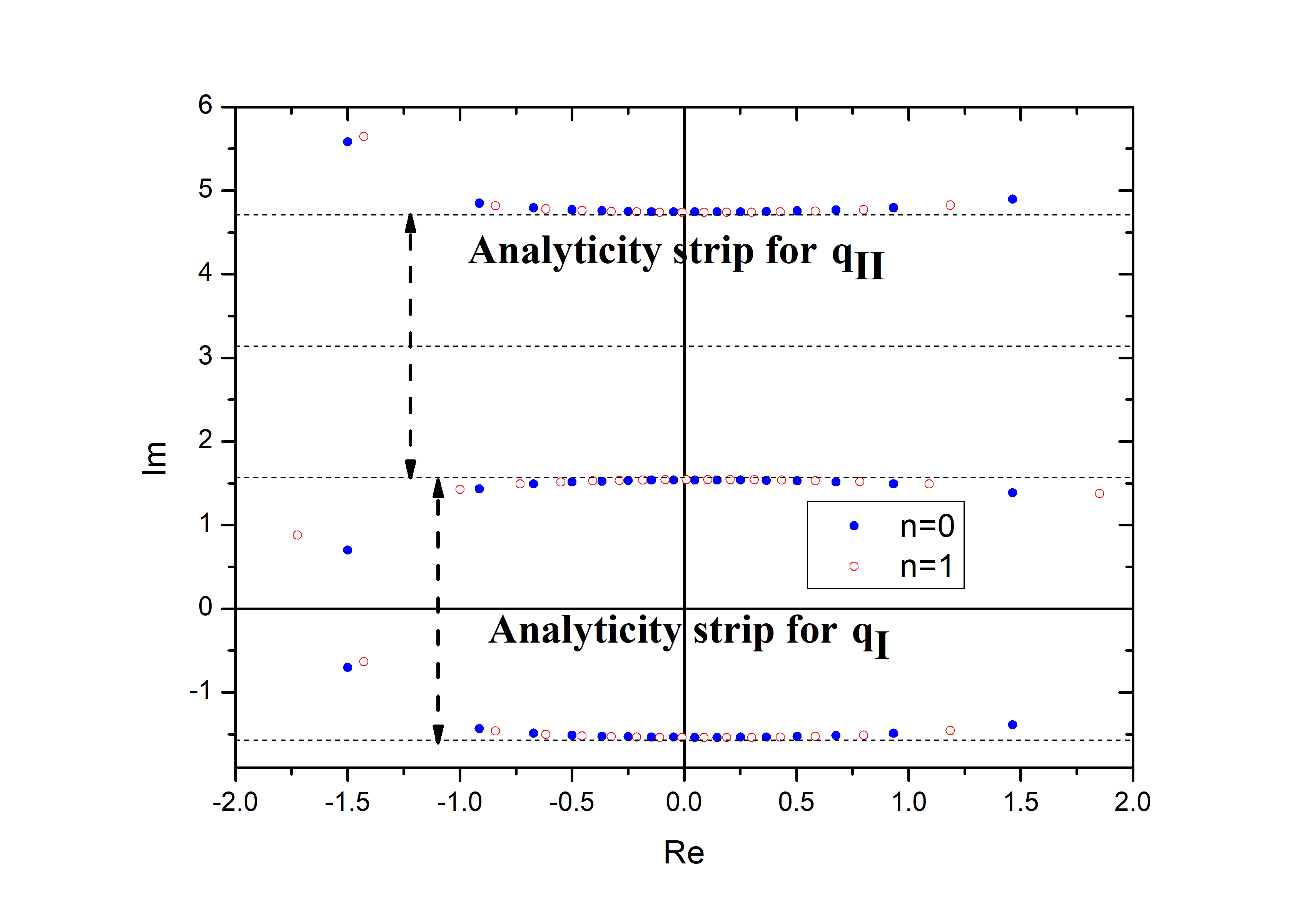}
\includegraphics[scale=0.3]{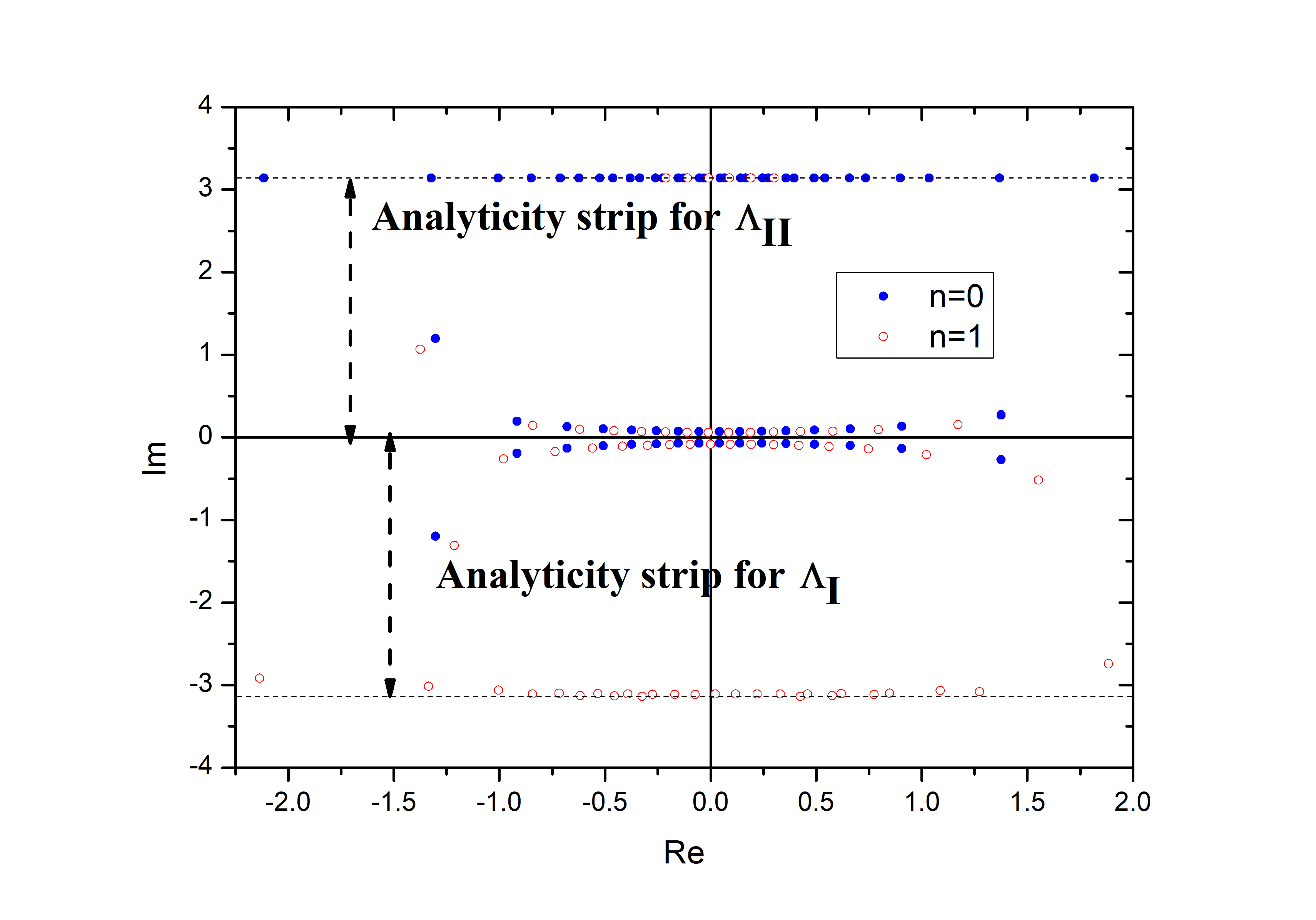}  
\caption{Descendant states with $N=1$ for $m=w=0$: Depiction of the location
  of all zeros in the complex plane for a) the function $q(z)$ and b) the
  eigenvalue function $\Lambda(z)$. We use parameters $L=32$ and
  $\gamma=0.8$. Disks refer to the lowest lying state in this sector ($n=0$)
  and circles to the first excited state ($n=1$). For these states the zeros
  of $q(z)$ and the zeros of $\Lambda(z)$ deviate noticeably from the standard
  distribution lines.  The zeros of $q(z)$ with strongest deviation (and
  largest negative real part) are denoted by $z_1, z_2$, the analogous zeros
  of $\Lambda(z)$ are denoted by $\thet_1, \thet_2$.  }
\label{fig:desc}
\end{center}
\end{figure}

\noindent
{\bf Descendant states}\\
Next we treat the first descendant state with $N=1$ for $m=w=0$. The
distribution of Bethe
roots, i.e.~the zeros of $q(z)$ are changed qualitatively. The two roots $z_1$,
$z_2$ with extremal real parts leave the axes with imaginary parts $\pm\pi/2$
and move closer to the real axis as illustrated in Fig.~\ref{fig:desc}. Also
the distribution of zeros of $\Lambda(z)$ changes. The two zeros with extremal
real parts leave the neighbourhood of the real axis as shown in Fig.~\ref{fig:desc}.
As above we define counter terms where we assume $\Im z_1>0$, $\Im z_2<0$,
$\Im \thet_1>0$, and $\Im \thet_2<0$
\begin{align}
  \lambda(z)&:=\frac{\sinh\tfrac12(z-\thet_1)\sinh\tfrac12(z-\thet_2)}
         {\sinh^2\tfrac12(z-\tfrac12(\thet_1+\theta_2))},\cr
         q_0(z)&:=\frac{\sinh\tfrac12(z-z_1)\sinh\tfrac12(z-z_2)}
         {\sinh\tfrac12(z-\Re z_1-\tfrac\pi2\ir)\sinh\tfrac12(z-\Re
           z_2+\tfrac\pi2\ir)},\cr
         \alpha(z)&:=\frac{q_0(z-2\ir\gamma)}{q_0(z+2\ir\gamma)},\qquad
         \beta(z):=\frac{q_0(z-\ir\gamma)}{q_0(z+\ir\gamma)}.
         \label{counterver3}
\end{align}
We introduce the modified functions
\begin{align}
  \tilde a_1(x)&:={\alpha(x+\ir\gamma)}a_1(x),
  \quad\tilde a_2(x):=\frac{a_2(x)}{\alpha(x+\ir\pi-\ir\gamma)},\cr
  \quad\tilde a_3(x)&:=\frac{a_3(x)}{\alpha(x-\ir\gamma)},
  \quad\tilde a_4(x):={\alpha(x+\ir\pi+\ir\gamma)}{a_4(x)},\label{tildeafunctionsdesc}
\end{align}
and
\begin{align}
  \tilde A_1(x)&:=\frac{\beta(x)}{\lambda(x+\ir\gamma)}A_1(x),
  \quad\tilde A_2(x):=\frac{A_2(x)}{\beta(x+\ir\pi)\lambda(x+\ir\pi-\ir\gamma)},\cr
  \quad\tilde A_3(x)&:=\frac{A_3(x)}{\beta(x)\lambda(x-\ir\gamma)},
  \quad\tilde A_4(x):=\frac{\beta(x+\ir\pi)}{\lambda(x+\ir\pi+\ir\gamma)}A_4(x).\label{tildeAfunctionsdesc}
\end{align}
These functions satisfy factorizations like
(\ref{deflittlea1})-(\ref{deflittlea4}) and (\ref{defbigA1})-(\ref{defbigA4}) with the
only change that $q(z)$ and $\Lambda(z)$ are replaced by
\begin{equation}
\tilde q(z):=\frac{q(z)}{q_0(z)},\qquad
\tilde\Lambda(z):=\frac{\Lambda(z)}{\lambda(z)},  
\end{equation}  
which have no zeros in the regions I and II as discussed above. Hence the
Fourier transforms of all equations yield (\ref{FTA1})-(\ref{FTA4}) where now
$\hat q_{I}, \hat q_{II}$ and
$\hToo$, $\hTot$ are the Fourier transforms of the logarithmic derivatives of
$\tilde q(z)$ and
$\tilde\Lambda(z)$ in regions I and II. From this again (\ref{singNLIE})
and (\ref{abbreviations}) follow with the only modification that $a_j$ and $A_j$ are to
be replaced by $\tilde a_j$ and $\tilde A_j$. The same holds for (\ref{regNLIE}) and
(\ref{finalNLIE}).

Now we can use (\ref{finalNLIE}) with the described replacements. Of course
the location of the $z_1, z_2$ and $\thet_1, \thet_2$ has to be known. They
are determined from the condition $a(z_j)=-1$ and $a(\thet_j)=-1$ or for
instance from $a_1(z_j-\ir\gamma)=-1$ and $a_1(\thet_j-\ir\gamma)=-1$ which are
evaluated by use of the NLIEs as integral expressions for the functions $\log
a_j(x)$ allowing for complex arguments $x$.

\begin{figure}
\begin{center}
  \includegraphics*[width=0.49\textwidth]{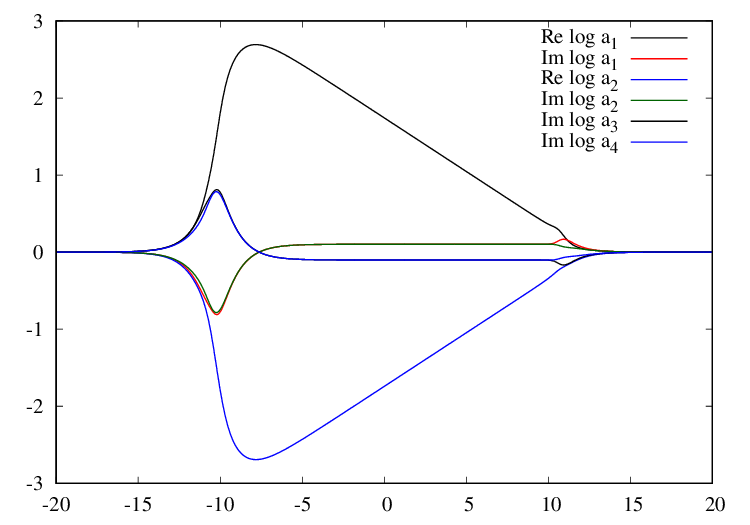}
  \includegraphics*[width=0.49\textwidth]{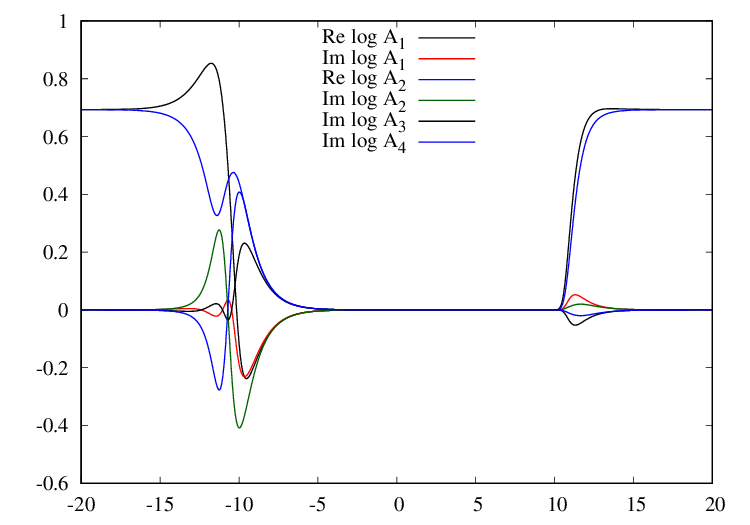}
  \caption{The first descendant state with $L/2-1$ roots located close to the
    upper as well as to the lower axis $\mathbb{R}\pm\ir\pi/2$ ($n=0$): a) Depiction of real and
    imaginary parts of the functions $\log \tilde a_j-d$ with $j=1, 2, 3, 4$,
    see (\ref{tildeafunctionsdesc}). The real parts for $i=1, 3$ ($i=2, 4$)
    are identical and shown by a black (blue) line.  Similar depiction of the
    functions $\log \tilde A_i$, see (\ref{tildeAfunctionsdesc}). We use
    parameters $L=10^{10}$ and $\gamma=0.9$.}
\label{FigDloga0}  
\end{center}
\end{figure}

\begin{figure}
\begin{center}
  \includegraphics*[width=0.49\textwidth]{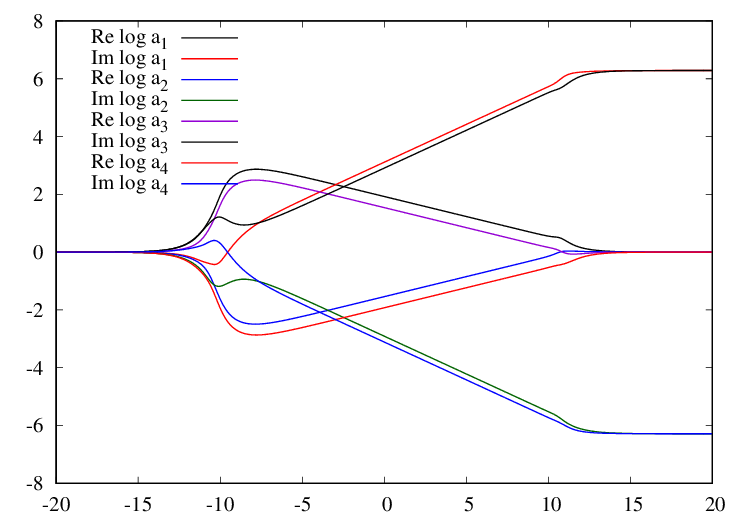}
  \includegraphics*[width=0.49\textwidth]{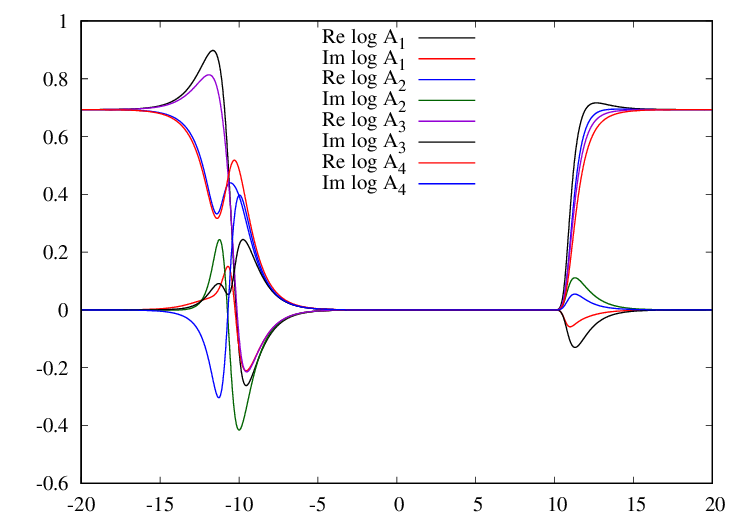}
  \caption{The second descendant state with $L/2$ roots located close to the
    upper and $L/2-2$ roots close to the lower axis ($n=1$): a) Depiction of real and imaginary
    parts of the functions $\log \tilde a_j-d$ with $j=1, 2, 3, 4$, see
    (\ref{tildeafunctionsdesc}).  Similar depiction of the functions $\log
    \tilde A_i$, see (\ref{tildeAfunctionsdesc}). We use parameters $L=10^{10}$
    and $\gamma=0.9$.}
\label{FigDloga1}  
\end{center}
\end{figure}

This program leads to a coupled set of NLIEs and scalar equations for the
$z_j$ and $\thet_j$. Having solved it, the integral expressions for energy and
quasi-momentum have to be evaluated. The modified versions of
(\ref{energyintegral}) and (\ref{qmomentum}) are
\begin{align}
  E&=L e_0
  +\frac{g\sin(2\gamma)}{\cosh\left(g\left(\thet_1-\frac\pi2\ir\right)\right)}
+\frac{g\sin(2\gamma)}{\cosh\left(g\left(\thet_2+\frac\pi2\ir\right)\right)}  
  -\frac{\sin(2\gamma)}{2\pi} \int_{-\infty}^\infty dx \,\frac{g^2 \cosh
    gx}{(\sinh gx)^2}\sum_{j=1}^4\log \Av_j(x) ,\label{energyintegralmoddesc}\\
  K&=\log\frac{\sinh(g(\thet_1-\ir\gamma))}{\sinh(g(\thet_2+\ir\gamma))}+
  \frac g{2\pi\ir}\int_{-\infty}^\infty dx \,\coth(gx)\,
\log\left(\frac{\Av_1(x)\Av_3(x)}{\Av_2(x)\Av_4(x)}\right).
\label{qmomentummoddesc}
\end{align}
In Fig.~\ref{FigDloga0} and Fig.~\ref{FigDloga1} we show the real and
imaginary parts of the above introduced functions for the descendant states
with $n=0$ and $n=1$.

\begin{figure}
\begin{center}
  \includegraphics[scale=0.5]{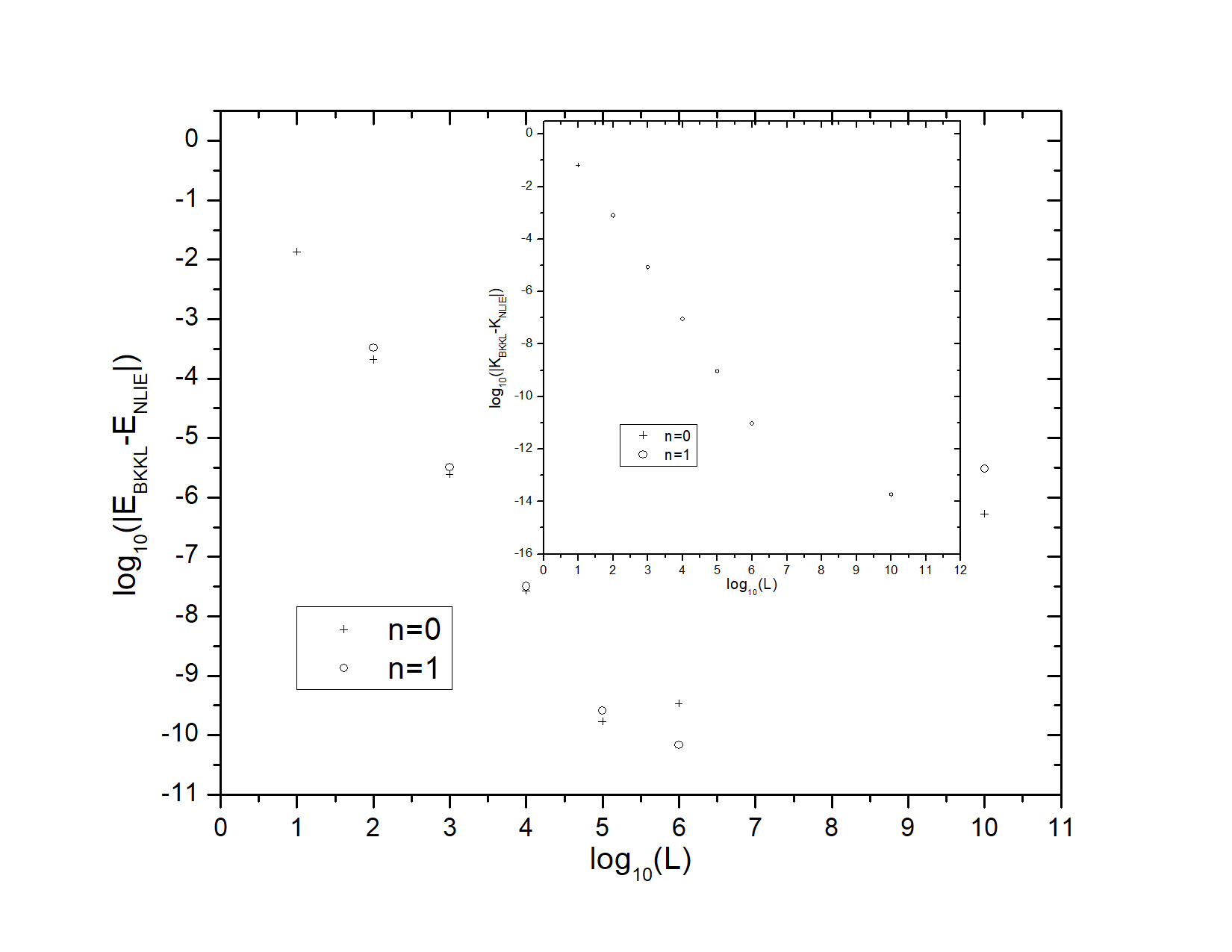}
  \caption{Comparison of the energies $E_{NLIE}$ and quasi-momenta $K_{NLIE}$
  for the descendant states with $E_{BKKL}=({2g\gamma}/{\pi})\,s^2$ and
  $K_{BKKL}=4g\gamma\,s$ with $s$ computed from the quantization condition in
  \cite{bazhanov2019scaling}. The parameters were taken to be
  $\gamma=0.9$ and $n=0, 1$. Plotted is the difference which vanishes algebraically
  with system size like $\mathcal{O}(L^{-2})$, but not below values of the order
  $10^{-10}$ ($10^{-14}$) for the energy (quasi-momentum) data.
  Our numerical calculations for the NLIEs were done  with double precision operations.
  The accuracy of the quasi-momentum data is higher than that for the energy
  data: The values of the quasi-momentum depend largely on the ``shape'' of
  the functions $\log A_j(x)$ as shown in Fig.~\ref{FigDloga0}  and
  Fig.~\ref{FigDloga1}. Note the transition of the functions between the asymptotic values 0
  and $\log(2)$ in regions $x\simeq\pm(\log L)/g$.
  The values of the energy depend sensitively on the precise location
  of this transition.
}
\label{fig:desacendants_comp}
\end{center}
\end{figure}

Fig.~\ref{fig:desacendants_comp} shows data obtained by our NLIE approach for
the descendant states compared to the ODE/IQFT results \cite{bazhanov2019scaling}.
We see the difference of the two approaches is best fitted by an order
$\mathcal{O}(L^{-2})$ ansatz. Here, for the descendant states with $N=1$ and $m=w=0$ the
quality of our numerical data is slightly worse than in the case of the
primary states with $N=0$ and $m=w=0$. This is understood from the numerics
and the ``richer features'' of the solution functions for the case $N=1$
rendering the numerical Fourier transforms less accurate. The form of the
graphs of the functions is the limiting factor, not the additional task to
determine the positions of the $\thet_j$ parameters. Their numbers appear to
be rather stable in various different runs of the iterative solution routine
with different parameters for the width of the interval and number of grid
points chosen. Note however, that for large system sizes the agreement
between the NLIE and the ODE/IQFT results for the quasi-momentum approaches 14
decimal digits. In case of the energy the agreement between the results
becomes worse for system sizes larger than $10^6$, see
Fig.~\ref{fig:desacendants_comp}. This we attribute to the more susceptible
dependence of (\ref{energyintegralmoddesc}) to numerical inaccuracies in
comparison to (\ref{qmomentummoddesc}).


\section{Conclusion}\label{sect:Conclusion}

In this paper, we have investigated the spectral properties of the staggered
six-vertex model with ${\cal Z}_2$ symmetry ($\alpha=\pi/2$) for arbitrary
system sizes $L$ using non-linear integral equations (NLIEs). Our study builds
on the important works of Candu and Ikhlef \cite{Candu_2013} and Frahm and
Seel \cite{Frahm14}, who were the first to address this topic and yielded
significant results in both the scaling limit and for finite, relatively large
system sizes.

Our research was motivated by two primary questions. The first question
concerned the accuracy of results based on the ODE/IQFT correspondence, which
holds in the asymptotic regime of large system sizes $L$. We found that the
quantization conditions for the low-lying primary states as obtained in
\cite{ikhlef2012conformal} and improved and extended to descendant states in
the comprehensive work of Bazhanov, Kotousov, Koval, and Lukyanov
\cite{bazhanov2019scaling,bazhanov2021scaling,bazhanov2021some}, are
impressively accurate even for relatively small sizes. In the anisotropy
parameter range $\pi/4 < \gamma < \pi/2$, the difference between NLIE and
ODE/IQFT results for the energy and quasi-momentum eigenvalues vanishes in the
scaling limit as $\mathcal{O}(L^{-2})$.

The second question we explored was whether there exists an optimal NLIE-based
approach for studying the staggered six-vertex model. The linear and
non-linear integral equations in \cite{ikhlef2012conformal} and
\cite{Candu_2013}, respectively, feature singular kernels, whereas those in
\cite{Frahm14} have regular kernels. We analyzed why this occurs
and how different versions of NLIE are related. We presented a compact
derivation of NLIE with a singular kernel, which is directly equivalent to
that in \cite{ikhlef2012conformal}. By rearranging terms, we obtained an
equivalent set of NLIE with a regular kernel. This version of NLIE still ``retains
some memory'' of the issues presented by the singular kernel version. While the
convolution integrals in the numerical treatment are well-defined, the
iterative procedure of the basic version of the regular kernel NLIE converges 
only for carefully designed initial data.

Both versions of NLIE, singuar and regular, are valid for all states with an
arbitrary number $n$ of Bethe roots reallocated between the distribution lines
$\mathbb{R} \pm \ir\pi/2$, where $n$ does not appear in the driving terms of
the NLIE. Naturally, the singular kernel NLIE introduces additional
challenges. In the regular kernel version, the kernel is not applied to the
four functions $\log A_j$, which take small values, but to the combination
$\log a_j -d- 2\log A_j$, which has non-vanishing asymptotes due to the winding
of the trajectories of the functions $a_j(x)$ in the complex plane. The value of the
windings is $\pm n$, meaning the imaginary parts of $\log a_j(x)$ increase by
$\pm n \cdot 2\pi$ between $x=-\infty$ and $x=+\infty$. By incorporating this
information, the regular kernel NLIE can be supplemented with counterterms to
allow for stable numerical solutions via iterative methods. Since all
functions entering the convolution integrals take small values, the numerical
results are highly accurate, up to nearly all available digits (with a minor
loss due to the large number of grid points).

We successfully performed calculations for lattice sizes ranging from $L=2$ to
$L=10^{24}$. For $L=2$ to $L=10^6$, the agreement with the ODE/IQFT results
increased, with deviations of the order of $\mathcal{O}(L^{-2})$. Beyond $L=10^6$, the
differences remained around $10^{-15}$ to $10^{-14}$ due to the ``limited"
numerical precision available. We assert that the results of the NLIE-based
calculations are numerically exact.

We also elucidated the relationship between our regular version of NLIE and
that of \cite{Frahm14}. A modified and slightly asymmetric variant
of our transformation from the singular kernel version to regular kernel
versions yields the results of \cite{Frahm14}. Finally, we
demonstrated how to derive analytic results for the scaling limit within the
NLIE approach. Interestingly, the singular version of NLIE is extremely useful
in this context, allowing for the application of the dilogarithmic trick
\cite{Klumper91}, which yields the leading logarithmic term in the conformal
weights. These calculations are conducted without any Wiener-Hopf techniques,
using instead only elementary manipulations and the winding of the
involved functions.

Many open questions remain, particularly regarding the derivation of all
higher-order terms identified by \cite{bazhanov2019scaling}.  Possibly similar
NLIE may allow for calculating the spectrum of the complex sinh-Gordon model
where in the lattice regularization the inhomogeneities contain a staggering
not only along the imaginary but also in the real direction.

\section*{Acknowledgments}
We acknowledge financial support by Deutsche Forschungsgemeinschaft through
FOR 2316. We thank Y.~Ikhlef, J.~Jacobsen, H.~Frahm, S.~Gehrmann, G.~Kotousov,
S.~Lukyanov, and H.~Saleur for stimulating discussions and insightful
perspectives on the topic.  AK gratefully acknowledges support from the Simons
Center for Geometry and Physics, Stony Brook University at which some of the
research for this paper was performed.  AK also gratefully acknowledges
support through the PIFI fellowship by the Chinese Academy of Sciences and the
Innovation Academy for Precision Measurement Science and Technology, Wuhan,
where the final calculations and writing were completed.

\newpage
\renewcommand{\thesection}{\Alph{section}}
\renewcommand{\theequation}{\thesection.\arabic{equation}}
\setcounter{section}{0} 
\setcounter{equation}{0} 

\section{Fourier transform}
By use of a short hand notation for the Fourier transform of the logarithmic
derivative of a function $f(x)$ as $\hat f:=\Ff_{k}\{\frac{d}{dx}\log f(x)\}$
we list the Fourier transforms of the logarithmic derivatives of the eight
multiplicative relations (\ref{deflittlea1})-(\ref{deflittlea4}) and
(\ref{defbigA1})-(\ref{defbigA4})
\begin{align}
\hat a_1&=\left( 1-{{\rm e}^{-2\gamma\,k}} \right) \hat\Phi+{{\rm e}^{-3\gamma\,k}}{\hat q_{II}}-{{\rm e}^{\gamma\,k}} \hat q_I,\\
\hat a_2&=\left( 1-{{\rm e}^{(2\gamma-\pi)\,k}}\right) \hat\Phi+{{\rm e}^{(3\gamma-\pi)\,k}}{\hat q_I}-{{\rm e}^{-(\gamma+\pi)\,k}}\hat q_{II},\\
\hat a_3&=\left( 1-{{\rm e}^{(2\gamma-\pi)\,k}}\right) \hat\Phi+{{\rm e}^{(3\gamma-2\pi)\,k}}\hat q_{II}-{{\rm e}^{-\gamma\,k}}{\hat q_I},\\
\hat a_4&=\left( 1-{{\rm e}^{-2\gamma\,k}} \right) \hat\Phi +{{\rm e}^{(\pi-3\gamma)\,k}}\hat q_I-{{\rm e}^{(\gamma-\pi)\,k}}\hat q_{II},
\end{align}
and
\begin{align}
\hat A_1&=-{{\rm e}^{-2\gamma\,k}}{\hat\Phi}+{{\rm e}^{-\gamma\,k}}{\hat q_I}-{{\rm e}^{\gamma\,k}} \hat q_I+{{\rm e}^{-\gamma\,k}}\hTot,\label{FTA1}\\
\hat A_2&=-{{\rm e}^{(2\gamma-\pi)\,k}}{\hat\Phi}+{{\rm e}^{(\gamma-\pi)\,k}}\hat q_{II}-{{\rm e}^{-(\gamma+\pi)\,k}}{\hat q_{II}}+{{\rm e}^{(\gamma-\pi)\,k}}{\hTot},\label{FTA2}\\
\hat A_3&=-{{\rm e}^{(2\gamma-\pi)\,k}}{\hat\Phi}+{{\rm e}^{\gamma\,k}} \hat q_I-{{\rm e}^{-\gamma\,k}}{\hat q_I}+{{\rm e}^{\gamma\,k}} \hToo,\label{FTA3}\\
\hat A_4&=-{{\rm e}^{-2\gamma\,k}}{\hat\Phi}+{{\rm e}^{-(\gamma+\pi)\,k}}{\hat q_{II}}-{{\rm e}^{(\gamma-\pi)\,k}}{ \hat q_{II}}+{{\rm e}^{(\pi-\gamma)\,k}}\hToo,\label{FTA4}
\end{align}
where we have explicitly
\begin{equation}
\hat\Phi=-\ir L{\frac {{\rm e}^{\pi \,k}}{{{\rm e}^{\pi \,k}}-1}}.
\end{equation}
These equations can be solved for $\hat a_1,..., \hat a_4$, $\hat q_{I}$, $\hat
q_{II}$, $\hToo$, $\hTot$ in terms of $\hat A_1,..., \hat A_4$, $\hat\Phi$.
The results for $\hat a_1,..., \hat a_4$ are given in
(\ref{singNLIE})-(\ref{block2}). Here we give the result for the Fourier
transform of the first logarithmic derivative of the function on the left
hand side of (\ref{LamEnerg}) and (\ref{defK})
\begin{align}
 {{\rm e}^{(\gamma-\pi)\,k}}\hTot+  {{\rm e}^{\gamma\,k}}\hToo&=
  -\ir L\frac{\cosh((2\gamma-\pi/2)k)}{\cosh((\pi/2-\gamma) k)\sinh((\pi/2)k)}
  {{\rm e}^{(\gamma-\pi/2)\,k}}
  +\frac{{{\rm e}^{(\gamma-\pi/2)\,k}}}{2\cosh((\pi/2-\gamma) k)}
  \left(\hat A_1+\hat A_2+\hat A_3+\hat A_4\right),\\
 {{\rm e}^{(\gamma-\pi)\,k}}\hTot-  {{\rm e}^{\gamma\,k}}\hToo &=
  \frac{{{\rm e}^{(\gamma-\pi/2)\,k}}}{2\sinh((\pi/2-\gamma) k)}
  \left(\hat A_1-\hat A_2+\hat A_3-\hat A_4\right).
\end{align}
The integrated inverse Fourier transform of this yields (\ref{LamEnerg}) and
(\ref{defK}).

The Fourier transform of the singular kernel needs additional ``definitions'' for the
treatment of the singularity. As it stands the kernel $K$ in
(\ref{FTKernelInv}) with (\ref{block1}) and (\ref{block2}) does not allow for
the (inverse) Fourier transform. Of course the second derivative of this
function, resp.~$K(k)$ multiplied by $-k^2$, can be Fourier transformed. Hence
all reasonable definitions of the Fourier transform will differ just by linear
terms in $x$ and constants. This freedom of definition does not imply that the
solution to the integral equation (\ref{singNLIE}) is lacking uniqueness. Of
course, the treatment requires care resp.~a reformulation as presented in
Sect.~\ref{sect:NLIEreg}.

We want to give the (inverse) Fourier transform of ${1}/\left[{2\sinh(\gamma
  k)\sinh((\pi-2\gamma)k)}\right]$, however for generic $\gamma$ this is difficult
and as intermediate goal we treat ${1}/\left[{2\sinh^2(\gamma
  k)}\right]$ by taking the $k$-integration contour ${\cal C}$ just below the
real axis. For the evaluation of the transform
we deform the contour into the upper half-plane to ${\cal
  C}+\ir\pi/\gamma$. Inbetween there is the pole of second order at $k=0$
which leads to an explicit contribution
\begin{equation}
I(k):=\int_{{\cal C}}dk\frac{\re^{\ir kx}}{2\sinh^2(\gamma  k)}=-\frac{\pi
  x}{\gamma^2}
+\re^{-\frac\pi\gamma x}\cdot I(k),
\end{equation}
and a remaining integral of the same form as on the left hand side,
because the shift of the contour goes into the exponential in the integrand
and yields a $k$-independent factor. From the last equation we get
\begin{equation}
\int_{{\cal C}}dk\frac{\re^{\ir kx}}{2\sinh^2(\gamma  k)}=-\frac{\pi
  x}{\gamma^2\left(1-\re^{-\frac\pi\gamma x}\right)},
\end{equation}  
with asymptotic behaviour $-(\pi/\gamma^2)\,x$ for large positive values of
$x$, and $0$ for large negative $x$.

With regard to the function of our interest we see from
\begin{equation}
  \frac{1}{2\sinh(\gamma k)\sinh((\pi-2\gamma)k)}
=\frac{g\gamma}{\pi}\frac{1}{2\sinh^2(\gamma k)}+\mathcal{O}(1/k),\quad g=\frac{\pi}{\pi-2\gamma},
\end{equation}
that the asymptotics of the (inverse) Fourier transform is
\begin{equation}
  -(g/\gamma)\,x \quad\hbox{for large positive}\ x,\qquad 0 \quad\hbox{for large negative}\ x,
\end{equation}
and possibly additive constants.
The next leading terms are of order $\exp(-(\pi/\gamma)|x|)$ and
  $\exp(-g |x|)$.\\

\section{ODE/IQFT results}
\setcounter{equation}{0} 
The ODE/IQFT analysis \cite{bazhanov2019scaling} yields the following quantization
condition for $s$. For arbitrary integer $n$
\begin{equation}
8s\
 \log\bigg(\frac{2L\,\Gamma(\frac{3}{2}+\frac{1}{\zeta})}{\sqrt{\pi}\,\Gamma(1+\frac{1}{\zeta})}\bigg)
 +\delta(s)-2\pi n=0\,,
\qquad
\left(\zeta:={\pi}/{\gamma}-2\right),\label{ODEIQFT1}
\end{equation}
where the 0 on the right hand side is actually an undetermined
$\mathcal{O}\big((\log L)^{-\infty}\big)$ term.  For the primary state $N=0$
with $m=w=0$, the phase shift $\delta$ entering the above equation is
explicitly given by the formula \cite{ikhlef2012conformal,bazhanov2019scaling}
\begin{equation}
  \delta(s)=\frac{16s}{n}\log(2)-4\log\left(2^{2s}\frac{\Gamma(\frac{1}{2}-{s})}
    {\Gamma(\frac{1}{2}+{s})}\right)\,.\label{ODEIQFT2}
\end{equation}
For the non-primary state with $N=1$, the explicit formulae for $\delta$ are
\begin{equation*}
\delta_\pm=-4\log\left(2^{\frac{2s}{\zeta}(\zeta+2)}
\frac{\Gamma(\frac{1}{2}-{s})}{\Gamma(\frac{1}{2}+{s})}
\frac{1-2s}{1+2s}\right)
-2\log\left(\frac{2\zeta \omega_\pm-(\zeta+2)(\zeta-2s)}{2n
\omega_\pm+(\zeta+2)(\zeta+2s)}\right)\,,\label{ODEIQFT3}
\end{equation*}
where
\begin{equation}
\omega_\pm=-\frac{\zeta+1}{2\zeta}\,\Bigg(2s\pm
\sqrt{{\zeta(\zeta+2)}-{4s^2}}\Bigg).\label{ODEIQFT4}
\end{equation}

\printbibliography

\end{document}